\documentclass[a4paper,11pt]{article}
\pdfoutput=1 

\usepackage{jheppub} 

\usepackage[T1]{fontenc} 
\usepackage{url}
\usepackage{amsmath}
\usepackage{graphicx,slashed,xcolor,multirow,bbold,mathtools,sidecap,tikz,bm,enumitem,booktabs,array}

\usepackage[normalem]{ulem}

\usepackage{tikz}
\usetikzlibrary{positioning,arrows}
\usetikzlibrary{decorations.pathmorphing}
\usetikzlibrary{decorations.markings}

\usepackage{subcaption}

\usepackage[compat=1.1.0]{tikz-feynman}
\newcommand{\beq}{\begin{equation}}
\newcommand{\eeq}{\end{equation}}
\newcommand{\bea}{\begin{eqnarray}}
\newcommand{\eea}{\end{eqnarray}}
\newcommand{\barr}{\begin{array}}
\newcommand{\earr}{\end{array}}

\usepackage{color}
\usepackage{colortbl}
\usepackage{cancel, slashed}

\long\def\/*#1*/{}
\usepackage{color}
 \usepackage[normalem]{ulem}
\definecolor{darkgreen}{cmyk}{1,0,1,0.4}
\definecolor{darkred}{cmyk}{0,1,1,0.4}

\title{\boldmath Revisiting the LHC Constraints on Gauge-Mediated Supersymmetry Breaking Scenarios }
\author[a]{Kirtiman Ghosh,}
\author[b]{Katri Huitu,}
\author[c]{Rameswar Sahu}

\affiliation[a,c]{\footnotesize Institute of Physics, Bhubaneswar, Sachivalaya Marg, Sainik School Post, Bhubaneswar 751005, India}
\affiliation[a,c]{\footnotesize Homi Bhabha National Institute, Training School Complex, Anushakti Nagar, Mumbai 400094, India}
\affiliation[b]{\footnotesize Department of Physics, and Helsinki Institute of Physics, University of Helsinki, Finland 00014}
\emailAdd{kirti.gh@gmail.com}
\emailAdd{katri.huitu@helsinki.fi}
\emailAdd{rameswar.s@iopb.res.in}

\abstract{Supersymmetry (SUSY) addresses several problems of the Standard Model, such as the naturalness problem and gauge coupling unification, and can provide cosmologically viable dark matter candidates. SUSY must be broken at high energy scales with mechanisms like gravity, anomaly, gauge mediation, etc. This paper revisits the Gauge Mediated SUSY Breaking (GMSB) scenarios in the context of data from the Large Hadron Collider (LHC) experiment. The ATLAS mono-photon search at 139 inverse femtobarn integrated luminosity at the 13 TeV LHC, in the context of a simplified General Gauge Mediation (GGM) scenario (which is a phenomenological version of GMSB with an agnostic approach to the nature of the hidden sector), relies on assumptions that do not hold across the entire parameter space. We identify a few crucial assumptions regarding the decay widths of SUSY particles into final states with gravitinos that affect the LHC limits on the masses of the SUSY particles. Our study aims to reinterpret the ATLAS constraints on the gluino-NLSP mass plane, considering all possible decay modes of SUSY particles in a realistic GGM model.

}

\begin{document} 

\tikzset{
  vector/.style={decorate, decoration={snake,amplitude=.4mm,segment length=2mm,post length=1mm}, draw},
  tes/.style={draw=black,postaction={decorate},
    decoration={snake,markings,mark=at position .55 with {\arrow[draw=black]{>}}}},
  provector/.style={decorate, decoration={snake,amplitude=2.5pt}, draw},
  antivector/.style={decorate, decoration={snake,amplitude=-2.5pt}, draw},
  fermion/.style={draw=black, postaction={decorate},decoration={markings,mark=at position .55 with {\arrow[draw=blue]{>}}}},
  fermionbar/.style={draw=black, postaction={decorate},
    decoration={markings,mark=at position .55 with {\arrow[draw=black]{<}}}},
  fermionnoarrow/.style={draw=black},
  scalar/.style={dashed,draw=black, postaction={decorate},decoration={markings,mark=at position .55 with {\arrow[draw=blue]{>}}}},
  scalarbar/.style={dashed,draw=black, postaction={decorate},decoration={marking,mark=at position .55 with {\arrow[draw=black]{<}}}},
  scalarnoarrow/.style={dashed,draw=black},
  electron/.style={draw=black, postaction={decorate},decoration={markings,mark=at position .55 with {\arrow[draw=black]{>}}}},
  bigvector/.style={decorate, decoration={snake,amplitude=4pt}, draw},
  particle/.style={thick,draw=blue, postaction={decorate},
    decoration={markings,mark=at position .5 with {\arrow[blue]{triangle 45}}}},
  gluon/.style={decorate, draw=black,
    decoration={coil,aspect=0.3,segment length=3pt,amplitude=3pt}}
}

\maketitle
\flushbottom

\section{Introduction}
\label{sec:intro}
Supersymmetry (SUSY) is one of the well-motivated and well-studied theories in high-energy physics (for interesting reviews, see \cite{Luty:2005sn, Bilal:2001nv, Weinberg:2000cr, Martin:1997ns}). SUSY provides solutions to many of the existing problems of the Standard Model (SM). To list a few, SUSY theories provide an effective solution to the naturalness problem \cite{Golfand:1971iw, Volkov:1973ix, Wess:1974tw}, it favors the unification of gauge couplings at some high energy scale \cite{Dimopoulos:1981wb, Dimopoulos:1981zb, Dimopoulos:1981yj, Sakai:1981gr, Ibanez:1981yh, Einhorn:1981sx}, and the particle spectrum of some of the R-parity conserving SUSY scenarios naturally includes a stable weakly interacting massive particle (WIMP) which can be a cosmologically viable candidate for dark matter \cite{Ellis:1983ew, Goldberg:1983nd, Mizuta:1992qp, Cirelli:2007xd, Feng:2000gh, Hisano:2006nn}. Despite all its successes, SUSY cannot be an exact symmetry of nature and must be broken (at least) softly \cite{Dimopoulos:1981zb, Girardello:1981wz, Sakai:1981gr} at some high energy scale. Several mechanisms, including gravity mediation \cite{Nilles:1982ik, Chamseddine:1982jx, Nath:1983aw, Barbieri:1982eh, Cremmer:1982vy, Ibanez:1982ee, Nilles:1982dy}, anomaly mediation \cite{Randall:1998uk, Giudice:1998xp}, gauge mediation \cite{Dine:1993yw, Dine:1994vc, Dine:1995ag}, etc., have been proposed in the literature as viable solutions to this problem. For our present analysis, we plan to look into the phenomenology of Gauge Mediated Supersymmetry Breaking (GMSB) scenarios in the context of the Large Hadron Collider (LHC) experiment. \\

The GMSB-type scenarios assume the SUSY breaking in the hidden sector is to be communicated to the visible sector through heavy messenger fields charged under the SM gauge groups. Since the gauge interactions are flavor-blind, GMSB provides a natural solution to the SUSY flavor problem and simultaneously avoids the stringent constraints of precision flavor observables. The minimal version of the GMSB scenario is highly predictive in the sense that the same messenger fields take part in the determination of the scalars and the gaugino spectrum resulting in a fixed ratio of the gluino, wino, and bino soft breaking masses ($M_3:M_2:M_1 = g_3^2:g_2^2:g_1^2$) at the messenger scale. Apart from the minimal GMSB scenario, there are many other possible realizations of gauge mediation (see reference \cite{Giudice:1998bp} for a review of such models). To facilitate the study of various realizations of the gauge mediation (including models with a strongly coupled hidden sector) in a manner that is agnostic to the nature of the hidden sector, Ref. \cite{Meade:2008wd, Buican:2008ws} proposed the framework of General Gauge Mediation (GGM). In the GGM construction, in the limit of vanishing gauge couplings, the MSSM decouples completely from the hidden sector responsible for the SUSY breaking. The GGM formalism is able to derive many features of the gauge mediation, including flavor universality, vanishing A-terms, gravitino LSP, and sfermion sum rules (see ref \cite{Meade:2008wd} for a detailed discussion). GMSB and its more general version, GGM, are some of the extensively studied frameworks in the SUSY literature.\\

Collider signatures of gauge mediation models with light gravitino in the context of colliders like LEP, Tevatron, and the LHC have been studied in Refs.~\cite{Ambrosanio:1996jn, Feng:1997zr, Stump:1996wd, Dimopoulos:1996va, Baer:2000pe, Kim:2011sv, Hiller:2009ii}. Studies of GGM scenarios in light of the Higgs mass constraints have been carried out in Refs. \cite{Knapen:2015qba,Knapen:2016exe}. The study of gauge mediation scenarios with electroweakino production at LHC has been thoroughly investigated in Refs. \cite{Aad:2015hea, ATLAS:2016hks, Kim:2017pvm}. Generation of large multi TeV A-terms (required for satisfying the Higgs mass in MSSM \footnote{Note that requiring a large A-term is not the only possible means to achieve a 125 GeV Higgs mass in the MSSM. The other possible solution is the existence of a very heavy stop. However, the latter solution is less attractive in the sense that such heavy stops cannot be accessed at current collider facilities.}\cite{Hall:2011aa, Heinemeyer:2011aa, Arbey:2011ab, Arbey:2011aa, Draper:2011aa, Carena:2011aa, Cao:2012fz,  Christensen:2012ei, Brummer:2012ns}) in gauge mediation scenarios has been studied in Refs. \cite{Evans:2011bea, Evans:2012hg, Kang:2012ra, Craig:2012xp, Abdullah:2012tq, Kim:2012vz, Byakti:2013ti, Craig:2013wga, Evans:2013kxa, Calibbi:2013mka, Jelinski:2013kta, Galon:2013jba, Fischler:2013tva, Knapen:2013zla, Ding:2013pya, Calibbi:2014yha, Basirnia:2015vga, Jelinski:2015gsa, Jelinski:2015voa}. Extension of gauge mediation models incorporating Higgs-messenger couplings has been studied in Refs. \cite{Komargodski:2008ax, Craig:2013wga}. For other related phenomenological studies on GGM, see Refs. \cite{Abel:2009ve, Abel:2010vba, Rajaraman:2009ga, Carpenter:2008he, Grajek:2013ola}. Our present paper is motivated by the ATLAS analysis \cite{ATLAS:2022ckd} that looked for the signature of a particular realization of the GGM scenario which is discussed briefly (see section \ref{sec:atlasanalysis} for a detailed discussion) in the following.\\

The ATLAS analysis \cite{ATLAS:2022ckd} looks into an R-parity conserving scenario. Therefore, SUSY particles can only be produced in pairs, and their decay must contain an odd number of SUSY particles in the final state. The parameter setting of the analysis is such that the gravitino is always the lightest SUSY particle (LSP) with the lightest neutralino as the next-to-lightest SUSY particle (NLSP). The ATLAS analysis \cite{ATLAS:2022ckd} is designed to search for the scenario where the NLSP neutralino is dominantly a mixture of bino and higgsino with negligible contribution from wino. Such an NLSP-neutralino dominantly decays into the LSP gravitino in association with a photon or a Z/Higgs boson. The pair production of gluinos (squarks are assumed to be decoupled at high energy) at the LHC results in a final state comprising of any combination of two bosons out of $\gamma$, Z, or Higgs (though the analysis always demands at least one high $p_T$ photon \footnote{The ATLAS analysis \cite{ATLAS:2022ckd} further concentrated on the region of parameter space that lead to a lightest neutralino which is an equal admixture of bino or higgsino. Such a NLSP decays to a gravitino-photon pair or gravitino-Z/H pair with almost equal branching fractions. Therefore, 75\% of the pair-produced gluino events lead to the final state with at least one hard photon.} in the signal region) resulting from the decay of NLSP neutralino to gravitino, many other SM particles (jets/leptons, W/Z/h boson, top-quark) resulting in the cascade decay of gluino to NLSP neutralino and a large missing transverse energy resulting from the two gravitinos in the final state. The analysis utilizes 139 $fb^{-1}$ integrated luminosity data collected by the ATLAS experiment at the $\sqrt{s} = 13$ TeV LHC. The non-observation of any significant excess in the data with respect to the SM prediction led the analysis to set lower bounds on the masses of gluino and NLSP neutralino of the GGM model considered.\\

Like most LHC analyses, the analysis in Ref.~\cite{ATLAS:2022ckd} is based on a simplified implementation of the GGM model with several assumptions that hold only in some particular regions of parameter space. For example, the assumption of equal branching ratios for the photon-gravitino and $Z$/Higgs-gravitino decay modes only holds for some specific values of soft breaking terms for bino and higgsino (see discussion in Section~\ref{sec:decay}), and the assumption of an eV-scale gravitino ensures the prompt decay of the NLSP-neutralino. Out of all these assumptions, we find one particular assumption, which has a significant impact on the final exclusion limits, cannot be satisfied over the entire region of parameter space under consideration. 

It is the assumption that except for the NLSP neutralino, the gravitino decay widths for the other SUSY particles are negligible compared to the decay width into lighter SUSY particles. In other words, decay into final states involving gravitino is not included for any other SUSY particle except for the NLSP neutralino. Consequently, the pair production of gluinos always results in a decay chain that ends up in the gravitino through the NLSP neutralino. We found that (detailed discussion in Section~\ref{sec:limitatlasanalysis}) this assumption does not hold for the entire gluino-neutralino mass considered in Ref.~\cite{ATLAS:2022ckd}. For those parts of the parameter space (where these assumptions do not hold), a proper reinterpretation of the experimental data as bounds on the gluino-NLSP mass plane is required, considering all possible decay modes for the SUSY particles. In our work, we will pursue this endeavor and recast the bounds on the gluino-NLSP mass plane in a realistic GGM model with all possible complexity of the decay of the SUSY particles.

The rest of the paper is organized as follows. The next section will introduce the GGM model and the SUSY particle spectrum, along with the decay of SUSY particles into gravitino final states. In section \ref{sec:atlasanalysis}, we will discuss the details of the ATLAS analysis \cite{ATLAS:2022ckd} followed by its limitations in section \ref{sec:limitatlasanalysis}. Next, we introduce the settings of our phenomenological analysis in section \ref{sec:analysis}. We present our final results in section \ref{sec:final}. Finally, we conclude our discussion in Section \ref{sec:summary}.

\section{Phenomenological Model}
\label{sec:model}
The model considered in our analysis is a simple extension of the MSSM with a light gravitino. Gravitinos are a natural prediction of local supersymmetric models. It is widely accepted that SUSY is broken in a hidden sector by a field singlet under the SM gauge transformations. The breaking of local SUSY gives rise to a massless spin-1/2 goldstino, which is then absorbed by the gravitino, giving it a mass $\approx \langle F \rangle/M_P$, where $\langle F \rangle$ is the vacuum expectation value (vev) associated with the F component of the hidden sector field responsible for breaking SUSY (see Ref. \cite{Ajaib:2012vc} for details), and $M_P$ is the Planck scale. SUSY breaking is then propagated to the visible sector by some messenger field. The SUSY literature provides various prescriptions for this propagation \cite{Randall:1998uk, Giudice:1998xp, Dine:1993yw, Dine:1994vc, Dine:1995ag, Nilles:1982ik, Chamseddine:1982jx, Nath:1983aw, Barbieri:1982eh, Cremmer:1982vy, Ibanez:1982ee, Nilles:1982dy}. For our analysis, we will concentrate on one such scenario, the Gauge Mediated SUSY Breaking (GMSB) \cite{Dine:1993yw, Dine:1994vc, Dine:1995ag} and its more general form, General Gauge Mediation (GGM) \cite{Meade:2008wd, Buican:2008ws}. In GMSB, SUSY breaking is mediated by messenger fields that interact with the visible and hidden sector fields through the usual SM gauge interactions. Since a gauge interaction is flavor blind, there is no SUSY flavor problem. By simple dimensional analysis, we can estimate the soft breaking masses to be of the order $\langle F \rangle/M_{mess}$, where $M_{mess}$ is the messenger scale \cite{Ajaib:2012vc}. Clearly, the gravitino mass experiences a natural suppression by a factor of $M_{mess}/M_P$ and is most likely the lightest SUSY particle (LSP) in GMSB-type models.

We are interested in the scenario where the lightest neutralino, a linear combination of the neutral higgsino ($\Tilde{H}^0_d,\Tilde{H}^0_u$) and gaugino $(\Tilde{B},\Tilde{W}^0)$ gauge eigenstates, is the next to lightest SUSY particle (NLSP). In the basis $\psi = (\Tilde{B},\Tilde{W}^0, \Tilde{H}^0_d,\Tilde{H}^0_u)^T$, the neutral gaugino and higgsino mass term has the form \cite{Martin:1997ns},
\begin{equation}
    \mathcal{L} = -\frac{1}{2} \psi^T M_N \psi,
\end{equation}
where,
\begin{equation}
    M_N = \begin{pmatrix}
            M_1 & 0 & -\frac{1}{2}g^{\prime}vc_{\beta} &\frac{1}{2}g^{\prime}vs_{\beta}\\
            0 & M_2 & \frac{1}{2}gvc_{\beta} & -\frac{1}{2}gvs_{\beta}\\
            -\frac{1}{2}g^{\prime}vc_{\beta}&\frac{1}{2}gvc_{\beta}&0&-\mu\\
            \frac{1}{2}g^{\prime}vs_{\beta}&-\frac{1}{2}gvs_{\beta}&-\mu&0
          \end{pmatrix}.
\label{eqn:neutralinomass}
\end{equation}
Here $M_1$, $M_2$, and $\mu$ are the bino, wino, and higgsino soft SUSY breaking parameters, $g$ and $g^{\prime}$ are the SM SU(2)$_L$ and U(1)$_Y$ gauge couplings, $v$ is the SM vev, $c_{\beta}$ and $s_{\beta}$ stands for ${\rm cos}\beta$ and ${\rm sin}\beta$, respectively, with ${\rm tan}\beta = v_u/v_d$, and $v_u$ and $v_d$ are the vevs associated with the up-type and down-type Higgs, respectively. The four physical neutralino mass eigenstates, $\Tilde{\chi}^0_i$ where $i=1, 2, 3, 4$, resulting from the diagonalization of neutralino mass matrix $M_N$ by an orthogonal diagonalising matrix N are given by,
\begin{equation}
    \Tilde{\chi}^0_i = N_{ij} \psi_j .
\label{eqn:neutralinomixing}
\end{equation}
The relative hierarchy among $M_1$, $M_2$, and $\mu$ determines the composition of $\Tilde{\chi}^0_i$ and its decay modes. We are particularly interested in the composition and decay of the lightest neutralino $\Tilde{\chi}^0_1$, which is the NLSP in our case. For $|M_1| \ll |M_2|, |\mu|$ the NLSP is bino-like, for $|M_2| \ll |M_1|, |\mu|$ the NLSP is wino like, and for $|\mu| \ll |M_1|, |M_1|$ the NLSP is higgsino like. Following the analysis in Ref. \cite{ATLAS:2022ckd}, we assume $|M_1|\sim |\mu| \ll |M_2|$, corresponding to an NLSP with an equal admixture of bino and higgsino components. Such NLSP decays to a photon + gravitino and Higgs + gravitino with almost equal branching ratio for $\mu < 0$. Similarly, for $\mu > 0$, it has an almost equal branching ratio for the photon + gravitino and $Z$ + gravitino decay modes (see next section for details).  Following Ref. \cite{ATLAS:2022ckd}, we have fixed the value of $M_2$ around 3 TeV and ${\rm tan}\beta$ at 1.5. $M_1 \approx \mu$ is varied to obtain the NLSP mass for scanning.\\
The charginos are linear combinations of the charged winos and higgsinos. In the basis of $\psi^{\pm} = (\Tilde{W}^+,\Tilde{H}^+_u,\Tilde{W}^-,\Tilde{H}^-_d)^T$ the bilinear term involving charged winos and higgsinos can be written as \cite{Martin:1997ns},
\begin{equation*}
    \mathcal{L} = -\frac{1}{2} (\psi^{\pm})^T M_C \psi^{\pm} + h.c,
\end{equation*}
where,
\begin{equation*}
    M_C = \begin{pmatrix}
           0 & X^T\\
           X & 0
          \end{pmatrix},
\end{equation*}
with
\begin{equation*}
    X = \begin{pmatrix}
        M_2 & \sqrt{2} s_{\beta}m_W\\
        \sqrt{2}c_{\beta}m_W & \mu
    \end{pmatrix}.
\end{equation*}
The mass matrix can be diagonalized by two $2 \times 2$ unitary matrices $U$ and $V$ such that 
\begin{equation*}
  U^*XV^{-1} = \begin{pmatrix} M_{\tilde{\chi}_1^{\pm}} & 0\\ 0 & M_{\tilde{\chi}_2^{\pm}} \end{pmatrix}
\end{equation*}
Following the approach of ref \cite{ATLAS:2022ckd}, we also assume the sleptons and squarks are decoupled at 5 TeV. Similarly, we have assumed all trilinear couplings to take a vanishing value and the Higgs sector apart from the SM Higgs to decouple at around 2 TeV. Next, we have the gluino soft breaking mass $M_3$, which determines the gluino mass, a free parameter for our analysis. Finally, we assumed a gravitino mass $\sim 1$ eV to allow prompt decay of the NLSP.\\
To summarize, the phenomenological model has two independent parameters $|M_1| \approx |\mu|$ and $|M_3|$. We have varied the free parameters $M_1$ and $M_3$ to obtain gluino and lightest neutralino mass in the following range for our final analysis: 1400 GeV $\leq M_{\Tilde{g}} \leq$ 2600 GeV and 150 GeV $\leq M_{\Tilde{\chi}_1^0} \leq$ 2600 GeV. In the next section, we discuss the possible decay modes of the SUSY particles that are relevant to our analysis.

\subsection{Decays}
\label{sec:decay}
The decay of supersymmetric particles is well-studied in the literature. Here, for completeness, we summarise the decay widths of the relevant SUSY particles into the gravitino final state. As discussed in the previous section, Ref. \cite{ATLAS:2022ckd} only considers the scenario where the lightest neutralino is an admixture of bino and higgsino gauge eigenstates. R-parity being conserved such a neutralino NLSP has three possible decay modes: $\gamma \Tilde{G}$, $Z \Tilde{G}$, and $h \Tilde{G}$. We present the neutralino partial decay widths into the $\gamma \Tilde{G}$, $Z \Tilde{G}$, and $h \Tilde{G}$ decay modes in the following \cite{Ambrosanio:1996jn}.
\begin{equation}
\begin{aligned}
    \Gamma(\Tilde{\chi}^0_i \rightarrow \gamma \Tilde{G}) &= \frac{k_{i\gamma}}{48 \pi} \frac{m_{\Tilde{\chi}^0_i}^5}{M_P^2 m_{\Tilde{G}}^2},\\
    \Gamma(\Tilde{\chi}^0_i \rightarrow Z \Tilde{G}) &= \frac{2k_{iZ_T}+k_{iZ_L}}{96 \pi} \frac{m_{\Tilde{\chi}^0_i}^5}{M_P^2 m_{\Tilde{G}}^2} \left(1-\frac{m_Z^2}{m_{\Tilde{\chi}^0_i}^2}\right)^4,\\
    \Gamma(\Tilde{\chi}^0_i \rightarrow h \Tilde{G}) &= \frac{k_{ih}}{96 \pi} \frac{m_{\Tilde{\chi}^0_i}^5}{M_P^2 m_{\Tilde{G}}^2} \left( 1-\frac{m_h^2}{m_{\Tilde{\chi}^0_i}^2} \right)^4,
\label{eqn:neutralinodecay}
\end{aligned}
\end{equation}
here, $k_{i\gamma} = |N_{i1} c_w + N_{i2} s_w|^2$, $k_{iZ_T} = |N_{i1} s_w - N_{i2} c_w|^2$, $k_{iZ_L} = |N_{i3} c_{\beta} - N_{i4} s_{\beta}|^2$, and $k_{ih} = |N_{i3} s_{\alpha} - N_{i4} c_{\alpha}|^2$. $N_{ij}$ are the elements of the neutralino mixing matrix defined in equation \ref{eqn:neutralinomixing}.\\

\begin{figure}[htb!]
	\centering
	\includegraphics[width=0.75\columnwidth]{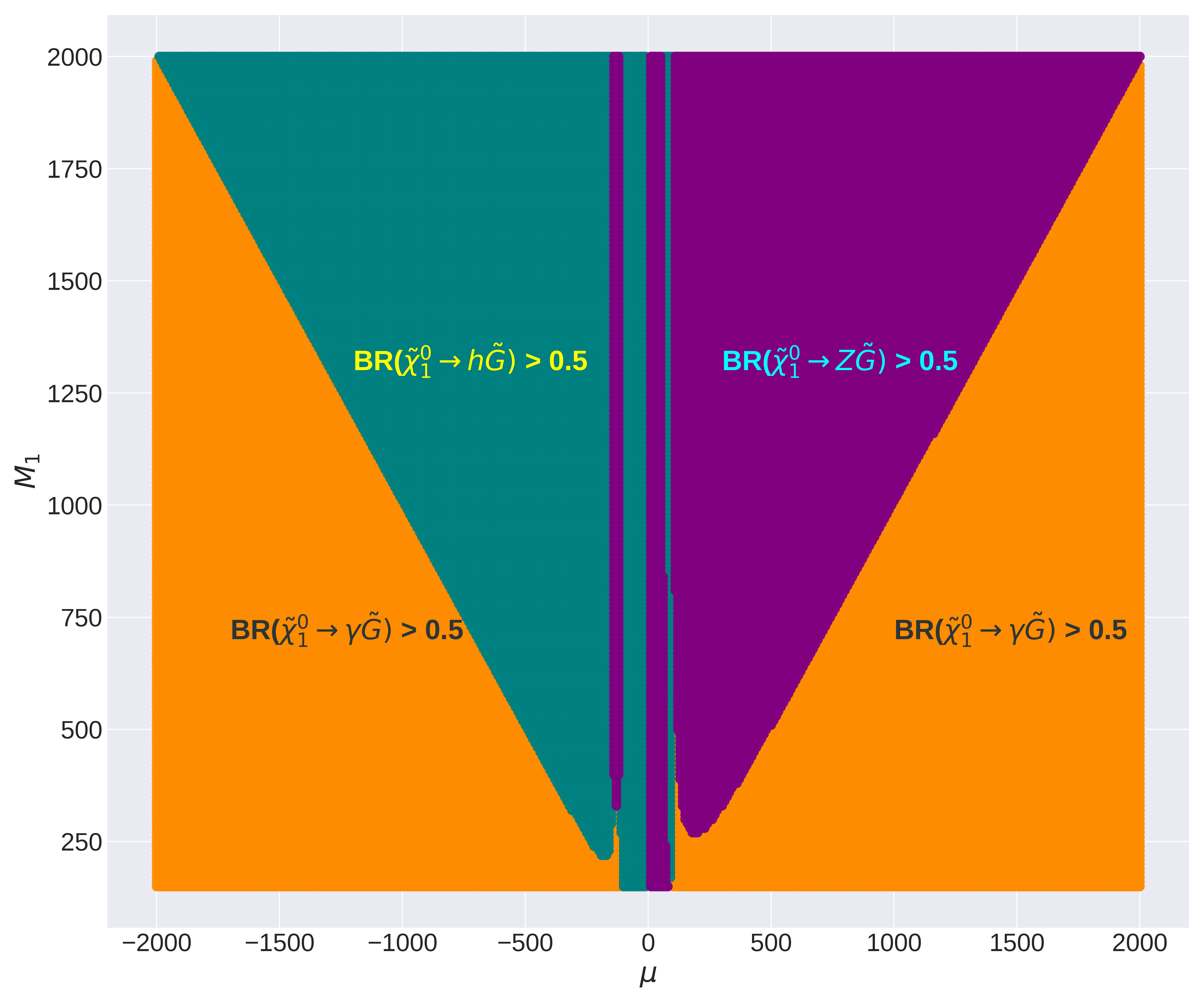} 
	\caption{\label{fig:Brnlsp1} Decay phase diagram of $\Tilde{\chi}_1^0$ segregating the $\mu$--$M_1$ parameter space with different decay modes' dominance. The region in orange, teal, and purple represent the parameter space where the decay branching ratio of the $\Tilde{\chi}_1^0$ into the final state with photon, Higgs, and Z boson exceeds 50\%, respectively.}
\end{figure}

\begin{figure}[htb!]
	\centering
	\includegraphics[width=0.45\columnwidth]{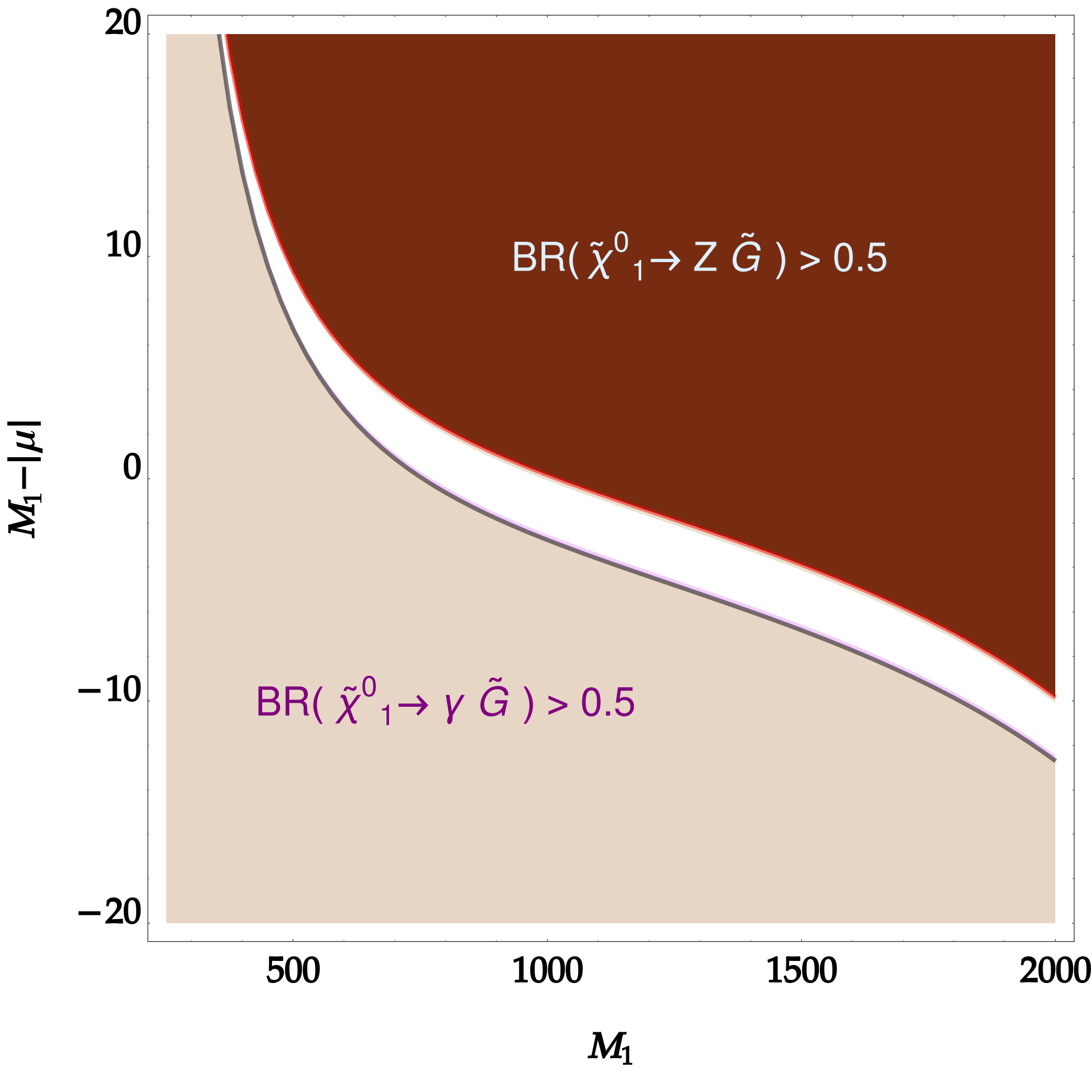} \qquad
	\includegraphics[width=0.45\columnwidth]{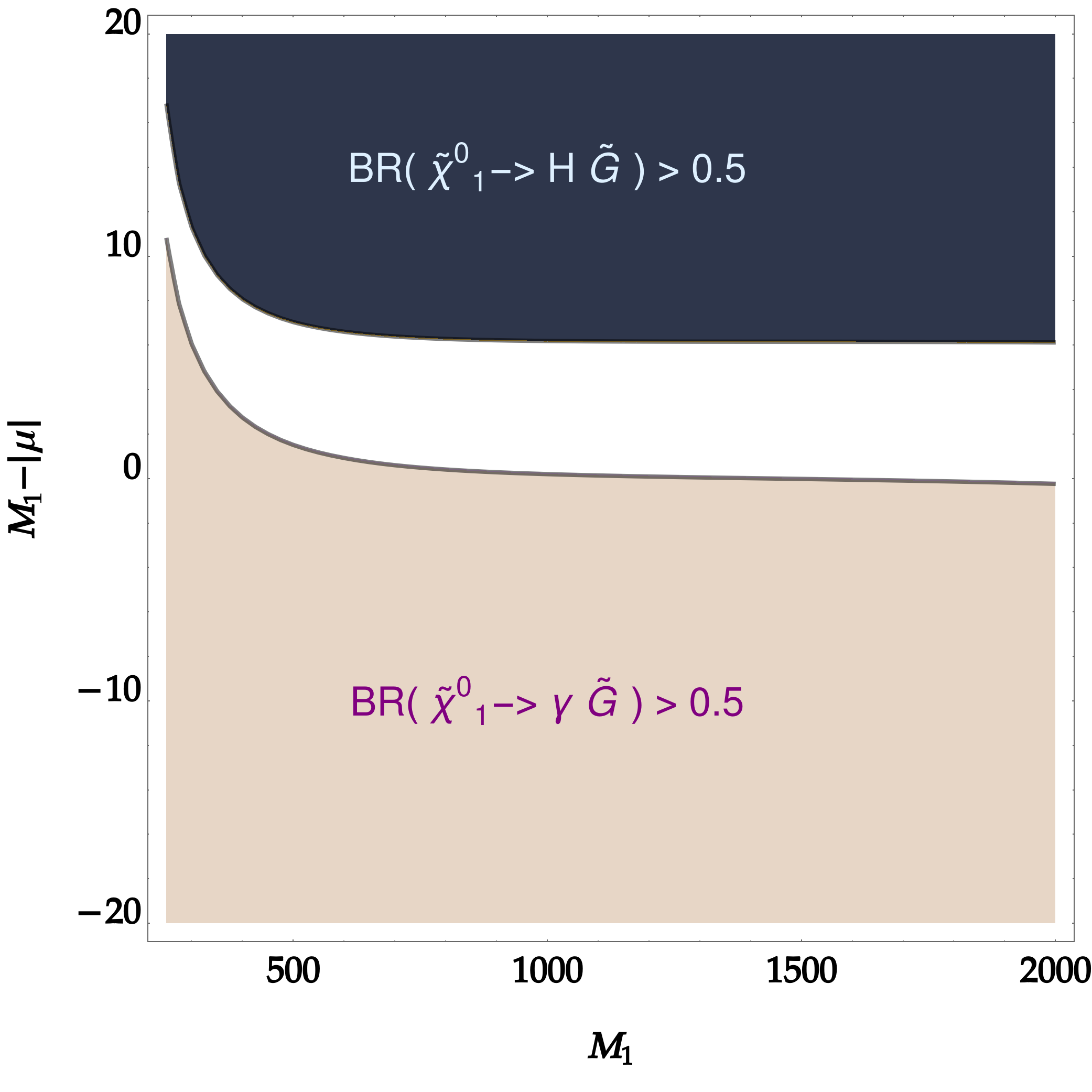} 
	\caption{\label{fig:Brnlsp2} Decay phase diagram of $\Tilde{\chi}_1^0$, for positive (left) and negative (right) values of $\mu$, segregating the $M_1$ Vs $(M_1-|\mu|)$ parameter space with different decay modes' dominance.  Pink/Orange solid contours correspond to 51\%, and black solid contours to 50\% branching ratios in different decay regions.}
\end{figure}

In Figure \ref{fig:Brnlsp1}, we present the decay phase diagram of the lightest neutralino (corresponding to $i=1$ in equation \ref{eqn:neutralinodecay}) for decay into these three decay channels on the $M_1$ vs $\mu$ plane. The ATLAS analysis \cite{ATLAS:2022ckd} is particularly interested in searching for the region of parameter space which gives rise to equal branching ratios for the neutralino NLSP into $\gamma~ \Tilde{G}$ and $Z/h ~ \Tilde{G}$ decay modes. Figure \ref{fig:Brnlsp1} shows that $|\mu|\approx M_1$ region of the $\mu$ vs. $M_1$ plane corresponds to such branching ratios for $\tilde{\chi}_1^0$. To obtain the exact region of the $\mu - M_1$ plane which results in 50 \% BR for both $\gamma~ \Tilde{G}$ and $Z/h ~ \Tilde{G}$ decay modes we present the decay phase diagram in Figure \ref{fig:Brnlsp2} on the $M_1-|\mu|$ vs $M_1$ plane for both positive (left) and negative (right) values of $\mu$. Note that for large values of $M_1$ and $M_1 \approx |\mu|$, the NLSP has almost equal branching ratios for the $\gamma + \Tilde{G}$ and h/Z + $\Tilde{G}$ channels (see Figure \ref{fig:Brnlsp2}). However, as we move towards smaller values of $M_1$, the photonic channel dominates because of phase space suppression in the other two decay modes. The black solid line in both plots of Fig. \ref{fig:Brnlsp2} corresponds to 50 \% branching fractions for $\tilde{\chi}_1^0 \rightarrow \gamma~ \Tilde{G}$ and $\tilde{\chi}_1^0 \rightarrow Z/h ~ \Tilde{G}$ decay modes. In the next part of our analysis, for a given $M_1$, the value of $\mu$ is given by the solid black lines to ensure equal branching ratios of the $\tilde{\chi}_1^0$ into $\gamma~ \Tilde{G}$ and $Z/h ~ \Tilde{G}$ decay modes. \\
The neutralinos $\Tilde{\chi}_2^0$, $\Tilde{\chi}_3^0$, and the chargino $\Tilde{\chi}_{1}^{\pm}$ are the next set of relevant particles that appear in the decay cascade of the gluino and hence impact the collider signature. The fourth neutralino and the second chargino are relatively heavier (see discussion in the previous section) and will not contribute to our analysis. The neutralino decay widths in the gravitino channel are already discussed in equation \ref{eqn:neutralinodecay}, and we present the chargino partial decay width into the $W^{\pm} \Tilde{G}$ decay mode in the following \cite{Ambrosanio:1996jn}:
\begin{equation}
    \Gamma(\Tilde{\chi}^+_i \rightarrow W^+ \Tilde{G}) = \frac{2k_{iW_T} + k_{iW_L}}{96 \pi} \frac{m_{\Tilde{\chi}^+_i}^5}{M_P^2 m_{\Tilde{G}}^2} \left( 1-\frac{m_w^2}{m_{\Tilde{\chi}^+_i}^2} \right)^4,
\label{eqn:charginonodecay}
\end{equation}
with,
\begin{align*}
    k_{iW_T} &= \frac{1}{2}(|V_{i1}|^2 + |U_{i1}|^2),\\
    k_{iW_L} &= |V_{i2}|^2 s_{\beta}^2 + |U_{i2}|^2 c_{\beta}^2,
\end{align*}
where, $i=1,2$ corresponding to $\Tilde{\chi}^{\pm}_1$ and $\Tilde{\chi}^{\pm}_2$. In addition to the gravitino decay modes discussed above, $\Tilde{\chi}_i^0$ ($i=2, 3$) and $\Tilde{\chi}_1^{\pm}$ can also decay into other SUSY particles when kinematically allowed. Such decay modes are referred to as cascade decay modes in the rest of the article. The parameter space studied in Ref. \cite{ATLAS:2022ckd} mostly favors three-body decays of these particles. For example, the second and third neutralinos ($\Tilde{\chi}_{2,3}^0$) or the lightest chargino ($\Tilde{\chi}_1^{\pm}$) undergo a tree-level three-body decay into a lighter neutralino or chargino in association with a pair of SM fermions. These decays proceed through an off-shell SM boson or slepton/squark in the propagator. Although due to the large mass of the slepton/squark, the latter contribution is subdominant compared to the contribution from the SM bosons. For a detailed discussion of these decay modes, see Ref. \cite{Djouadi:2001fa}. The relative strength of the gravitino decay\footnote{Note that in Ref.~\cite{ATLAS:2022ckd} the gravitino decays for $\Tilde{\chi}_i^0$ ($i=2, 3$) and $\Tilde{\chi}_1^{\pm}$ are assumed to be negligible compared to the cascade decays and hence ignored.} and the cascade decays for $\Tilde{\chi}_i^0$ ($i=2, 3$) and $\Tilde{\chi}_1^{\pm}$ are crucial to determine the final state signatures resulting from the gluino pair production at the LHC, which is discussed in the following.

\begin{figure}[htb!]
	\centering
	\includegraphics[width=0.3\columnwidth]{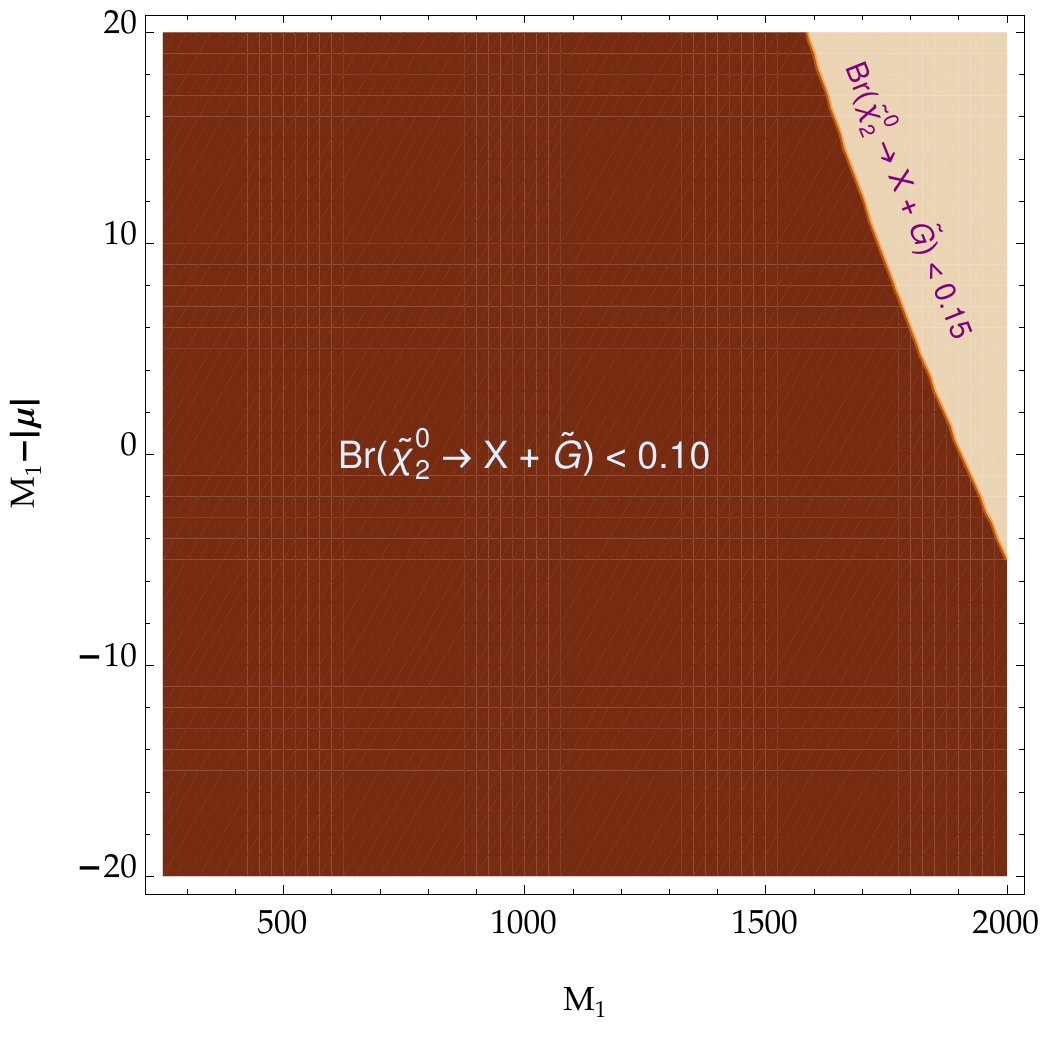} 
	\includegraphics[width=0.3\columnwidth]{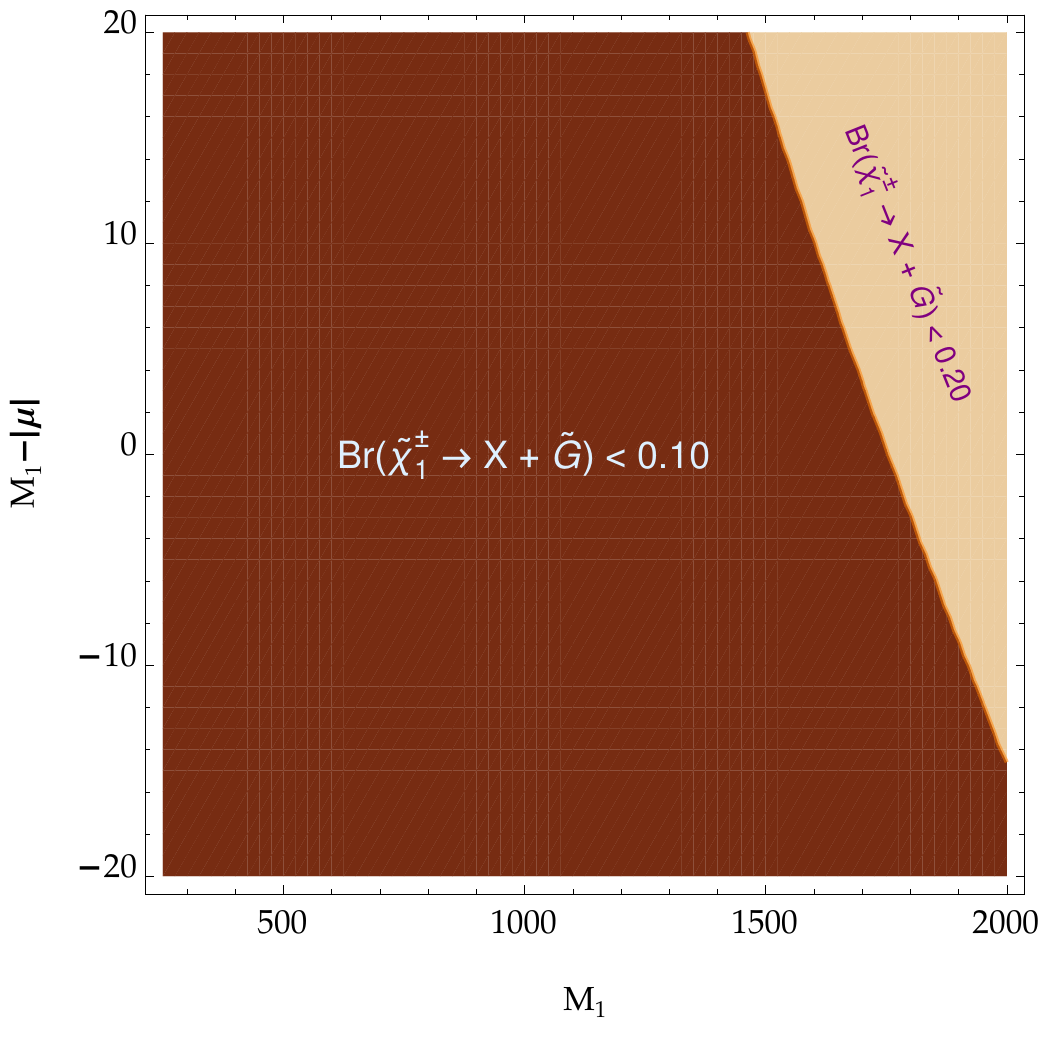} 
	\includegraphics[width=0.3\columnwidth]{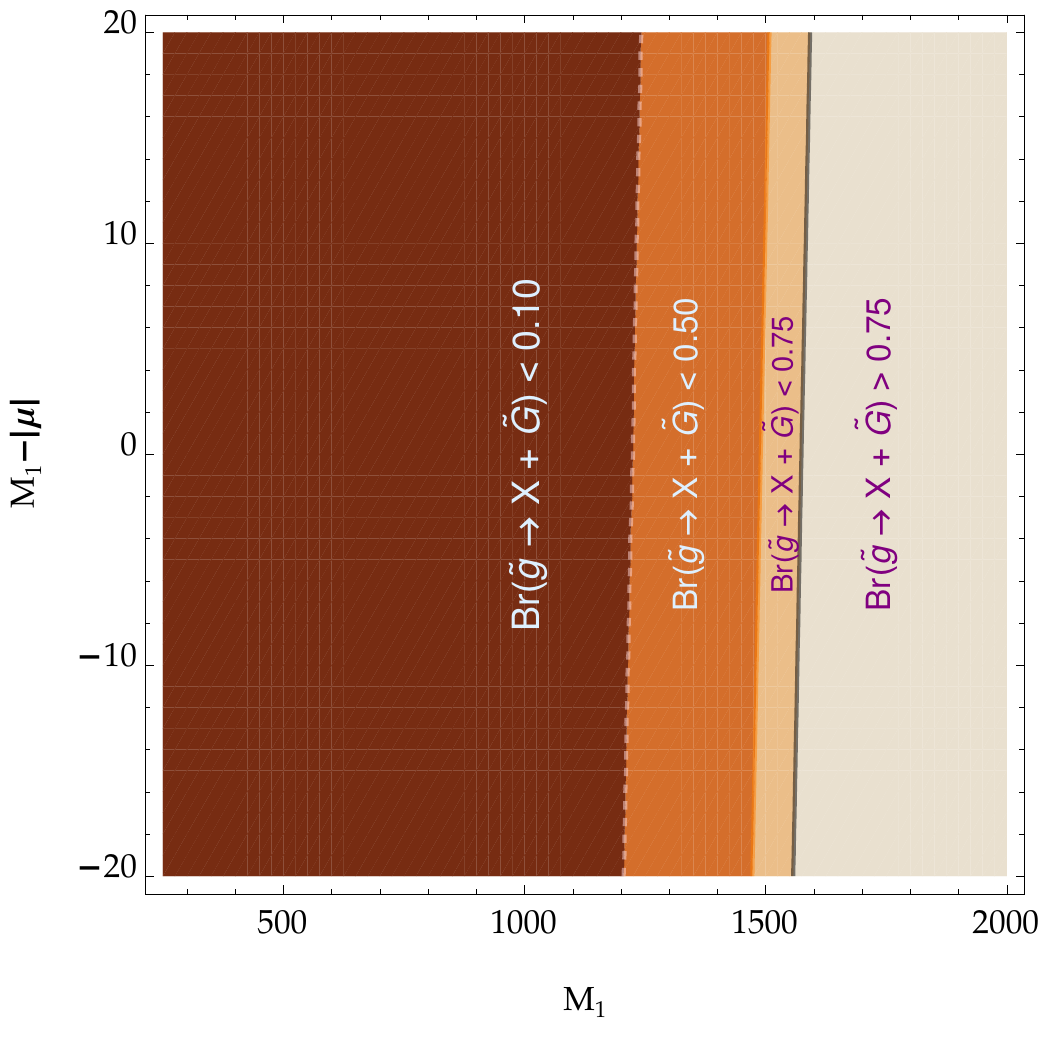}
	\caption{\label{fig:Brmupos} Decay phase diagram of $\Tilde{\chi}_2^0$(left), $\Tilde{\chi}_1^{\pm}$(middle), and $\tilde{g}$(right), for positive values of $\mu$ segregating the $M_1$ Vs $(M_1-|\mu|)$ parameter space, showcasing the branching ratios in the gravitino Decay channel.}
\end{figure}

\begin{figure}[htb!]
	\centering
	\includegraphics[width=0.3\columnwidth]{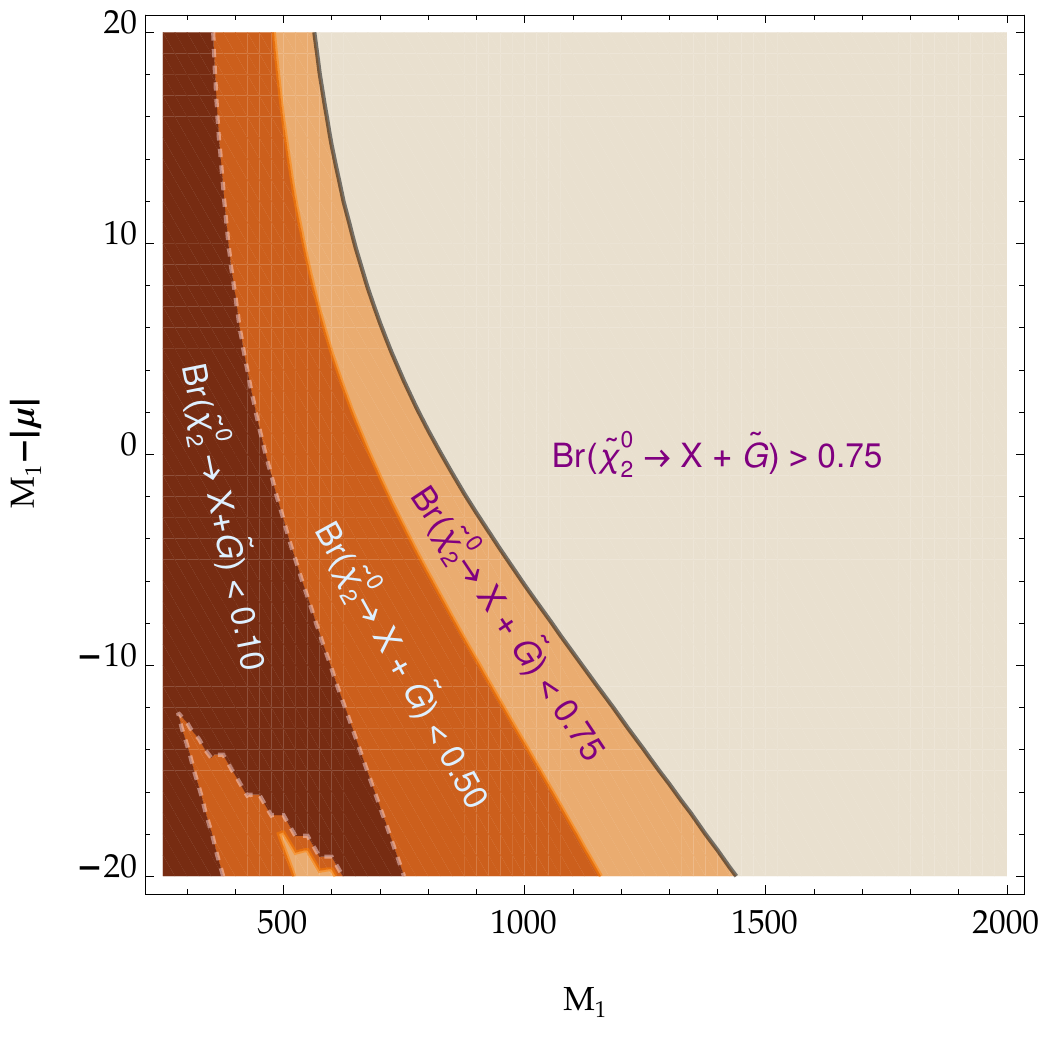} 
	\includegraphics[width=0.3\columnwidth]{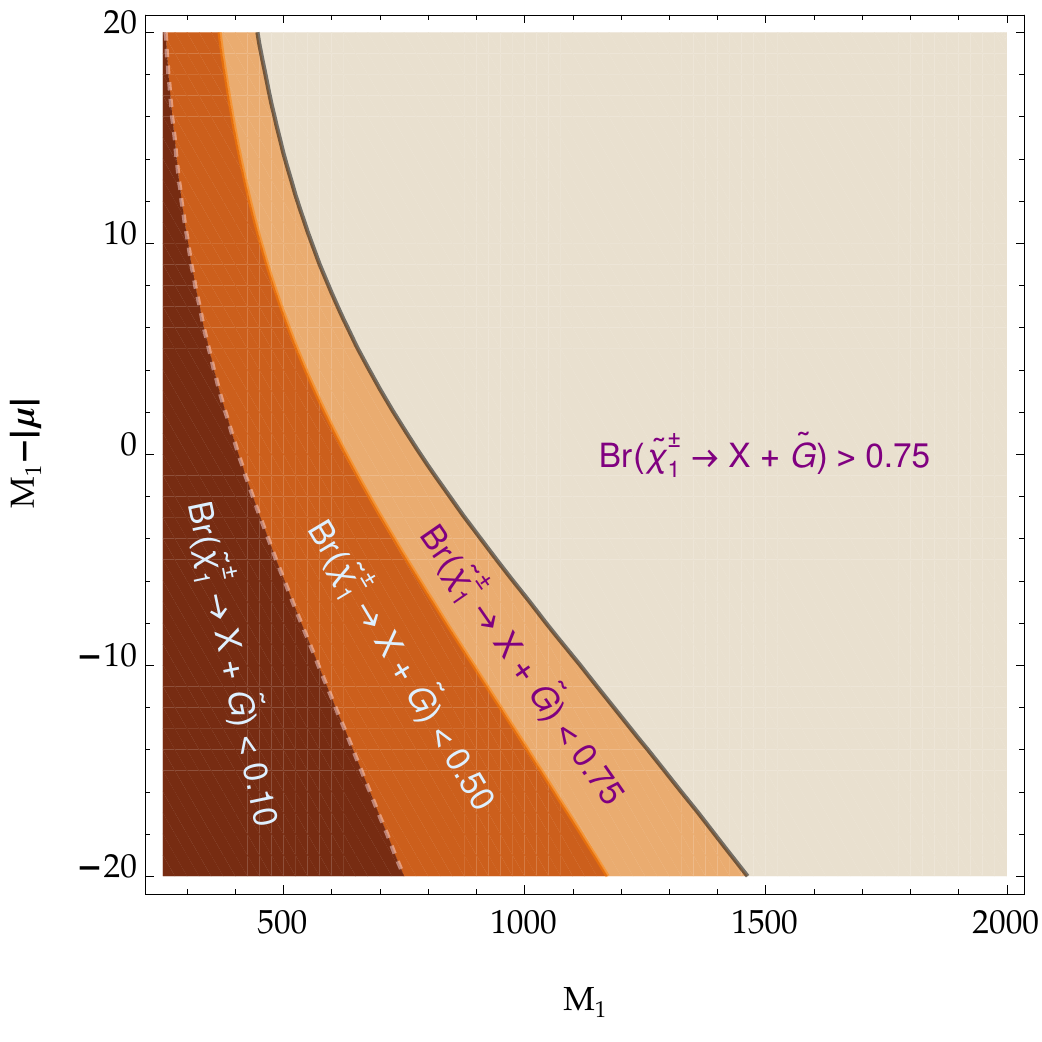} 
	\caption{\label{fig:Brmuneg} Decay phase diagram of $\Tilde{\chi}_2^0$(left), and $\Tilde{\chi}_1^{\pm}$(right), for negative values of $\mu$ segregating the $M_1$ Vs $(M_1-|\mu|)$ parameter space, showcasing the branching ratios in the gravitino Decay channel.}
\end{figure}

\begin{figure}
    \centering
    \includegraphics[width=0.8\columnwidth]{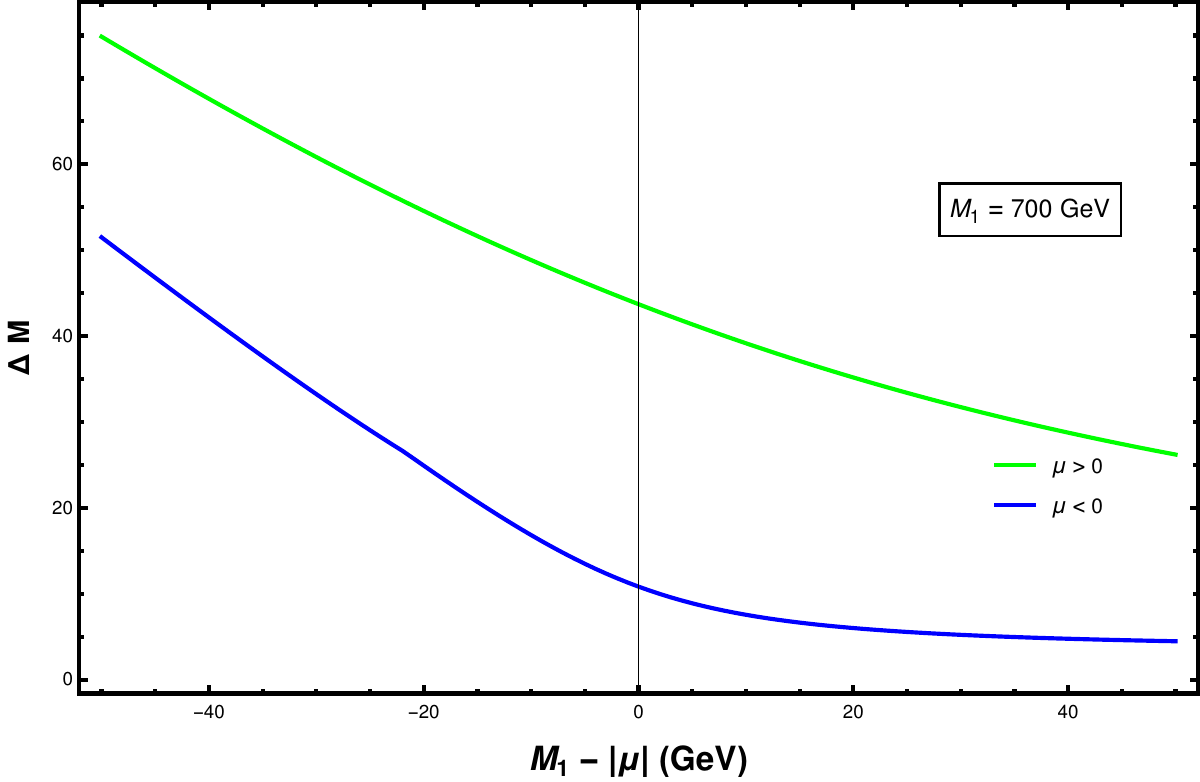}
    \caption{Variation of the mass difference between the Next-to-lightest neutralino and the lightest neutralino with $M_1 - |\mu|$ for positive and negative values of $\mu$}
    \label{fig:mass-diff}
\end{figure}

In the left and middle panel of Figure \ref{fig:Brmupos}, we present the decay phase diagrams of $\Tilde{\chi}_2^0$ and $\Tilde{\chi}_1^{\pm}$, respectively, for positive values of $\mu$, in the $M_1$ vs $M_1-|\mu|$ plane. Similar results for negative values of $\mu$ are presented in the left and middle panels of Figure \ref{fig:Brmuneg}. For the purpose of discussion, we have categorized the decay modes of these particles into two groups. The first group includes the decay mode with a final state gravitino (called gravitino decay mode in the following), while the other group includes all other possible decay modes, which we refer to as the cascade decay mode in the following discussion. The numbers in the plot represent the branching ratios in the gravitino decay mode. Figure~\ref{fig:Brmupos} for $\mu > 0$ shows that the cascade decays for the $\Tilde{\chi}_2^0$ (left panel) and $\Tilde{\chi}_1^{\pm}$ (middle panel) remain the dominant decay modes for the entire region of the $M_1$--$|\mu|$ plane of our interest (see Fig.~\ref{fig:Brnlsp2}), thereby validating the ATLAS assumption of negligible gravitino decays for $\Tilde{\chi}_2^0$ and $\Tilde{\chi}_1^{\pm}$. On the other hand, in Figure~\ref{fig:Brmuneg} for $\mu < 0$, gravitino decays remain significant for $\Tilde{\chi}_2^0$ (left panel) and $\Tilde{\chi}_1^{\pm}$ (middle panel) in a large part of the $M_1$--$|\mu|$ plane and hence cannot be ignored. For a qualitative understanding of the results, we also present the mass splitting between the next-to-lightest neutralino and the lightest neutralino (i.e., $\Delta M (\Tilde{\chi}_2^0,\Tilde{\chi}_1^0)$) as a function of $M_1 - |\mu|$ for both positive and negative values of $\mu$ in Figure \ref{fig:mass-diff}. The lightest chargino mass exhibits a similar behavior, and we refrain from presenting the results for brevity. From Figure \ref{fig:mass-diff}, we can make the following two observations,
\begin{itemize}
    \item $\Delta M(\Tilde{\chi}_2^0,\Tilde{\chi}_1^0)$ reduces gradually with $M_1-|\mu|$ in both scenarios, resulting in greater phase space suppression for the cascade decay channels.
    \item The phase space suppression in the $\mu < 0$ scenario is comparatively higher. As a result, for the $\mu < 0$ scenario, the gravitino decay mode can easily compete with the cascade decay modes for most parts of the parameter space shown in Figure \ref{fig:Brmuneg}.
\end{itemize}
These two factors combined can beautifully explain the behavior of the $\Tilde{\chi}_2^0$ and $\Tilde{\chi}_1^{\pm}$ decay phase diagrams shown in Figure \ref{fig:Brmupos} for the $\mu > 0$ scenario and the behavior of $\Tilde{\chi}_1^{\pm}$ decay phase diagram in Figure \ref{fig:Brmuneg} for the $\mu < 0$ scenario. \\


The partial decay width of gluino into gravitino LSP is given by \cite{Ambrosanio:1996jn}:
\begin{equation}
    \Gamma(\Tilde{g} \rightarrow g \Tilde{G}) = \frac{m_{\Tilde{g}}^5}{48 \pi M_P^2 m_{\Tilde{G}}^2}
\label{eqn:gluinodecay}
\end{equation}
In Figure \ref{fig:Brmupos}, we demonstrate the variation of the branching ratio of the gluino in the cascade decay channel in the $M_1-|\mu|$ Vs $M_1$ plane for positive $\mu$, for a fixed value of $M_{\Tilde{g}} = $ 2000 GeV. We omit the results for negative $\mu$ values, as the gluino decay width shows only a weak dependence on the sign of $\mu$. From equation  \ref{eqn:gluinodecay}, it is clear that the decay width into gravitino does not depend on the value of $M_1$ and $\mu$, and for the choices of parameters considered here, it has a value of $3.684 \times 10^{-5}$ GeV. The same is not true for the cascade decay channel. The main processes contributing to the cascade decay are the loop-induced two-body decays $\Tilde{g}\rightarrow g \Tilde{\chi}^0_i$, $i$=1,2,3 and the three-body decays  $\Tilde{g}\rightarrow q \Bar{q} \Tilde{\chi}^0_i$ ($i$=1,2,3) and $\Tilde{g}\rightarrow q \Bar{q^{\prime}} \Tilde{\chi}^{\pm}_1$ (see Ref. \cite{Barbieri:1987ed} for a detailed discussion). All these decay modes depend strongly on the mass difference between the gluino and the neutralino/chargino. With increasing $M_1$, this mass difference reduces, resulting in greater phase space suppression and reduced cascade decay width. We observe this behavior clearly in the right panel of Figure \ref{fig:Brmupos}.

\section{Collider phenomenology}
\label{sec:phenomenology}
\subsection{Brief Summary of the ATLAS search in Ref. \cite{ATLAS:2022ckd}}
\label{sec:atlasanalysis}
In this section, we will provide a brief review of the ATLAS analysis \cite{ATLAS:2022ckd}. The analysis focuses on the GGM-type SUSY breaking scenario with an ultralight gravitino. The details of the model, parameter space, particle spectrum, decays of relevant SUSY particles, etc., are already discussed in Section \ref{sec:model}. In the models studied in Ref. \cite{ATLAS:2022ckd}, the squarks are relatively heavy and are out of the LHC reach. Therefore, gluino pair production takes over as the dominant production mode of the SUSY particles due to its associated color charge. Also, Ref. \cite{ATLAS:2022ckd} considers SUSY models that conserve R-parity. Therefore, each SUSY particle can decay into a final state with an odd number of SUSY particles. On top of this, the experimental analysis \cite{ATLAS:2022ckd} imposes an additional condition that only allows the NLSP neutralino to decay into final states with gravitino. In other words, it assumes the gravitino decay mode of all SUSY particles, except for the NLSP neutralino, is suppressed compared to the other allowed decay modes (more on this in the next section). As a result, at LHC, the produced gluinos always generate a decay cascade that passes through SUSY particles of intermediate masses and terminates at the LSP (the gravitino), resulting in missing transverse momentum signatures. The decay of intermediate particles in the decay chain produces many quarks and leptons, though the leptonic final state is not considered in the analysis. The decay to the LSP gravitino always proceeds through the NLSP neutralino, which decays dominantly into the $\gamma$ + gravitino and $Z/h$ + gravitino channels with equal branching ratios (for details, see the discussion in Section \ref{sec:decay}). Therefore, the pair production of gluinos in the context of the particular region of GGM parameter space and assumptions of Ref. \cite{ATLAS:2022ckd} gives rise to multiple hard photons, $Z/h$ bosons in association with a number of jets/leptons and large missing transverse energy signature at the LHC. The ATLAS analysis in Ref. \cite{ATLAS:2022ckd} considered a final state with at least one high $p_T$ photon, many jets, and a large missing transverse momentum to search for the signatures of gluino production in the framework of simplified GGM scenario described above.\\

To define the signal, validation, and control regions, photon, lepton, and jet candidates (called signal-region candidates in Ref. \cite{ATLAS:2022ckd}) are selected based on the following requirements (see Ref. \cite{ATLAS:2022ckd} for a detailed discussion): 
\begin{enumerate}
    \item Photon candidates are required to have $p_T > 50$ GeV, $|\eta| < 2.37$, and must lie outside the electromagnetic calorimeter (ECAL) barrel-endcap transition region (1.37 < $|\eta|$ < 1.52).
    \item Additional conditions on the calorimetric isolation energy ($E_T^{\rm iso}$, defined as the sum of transverse energies of the topological clusters within $\Delta R = 0.4$ of the cluster barycentre) and track isolation variable ($p_T^{\rm iso}$, computed as the scalar sum of the transverse momentum of good-quality tracks ($p_T > 0.5$ GeV) within $\Delta R = 0.2$ of the photon candidate) are imposed to remove background events where jets are misidentified as photons. The analysis demands $E_T^{\rm iso} < 2.45~ {\rm GeV} + 0.022~ p_T$ and $p_T^{\rm iso} < 0.05~ p_T$ , where $p_T$ is the transverse momentum of the photon.
    \item Electron candidates are required to have $p_T > 25$ GeV, $|\eta| <$ 2.47, satisfy a 'loose' isolation requirement \cite{ATLAS:2019jvq}, and are removed if they are within the ECAL barrel-endcap transition region. 
    \item Muon candidates are required to have $p_T > 25$ GeV, $|\eta| <$ 2.7, and satisfy a 'loose' isolation criteria \cite{ATLAS:2020auj}.
    \item Jets are reconstructed using the anti-$k_t$ jet clustering algorithm \cite{Cacciari:2008gp, Cacciari:2011ma} with a radius parameter $R=0.4$ and are required to satisfy $p_T > 30$ GeV and $|\eta| < 2.5$ 
    \item Missing transverse momentum is computed from the negative vector sum of the transverse momentum of the reconstructed objects and tracks not associated with these objects.
    \item To avoid possible double counting of selected objects, a procedure based on Ref. \cite{ATLAS:2016poa, ATLAS:2020uiq} is used. 
\end{enumerate}
The difference between $M_{\Tilde{g}}$ and $M_{\Tilde{\chi_1^0}}$ determines the kinematics of the final state particles in the signal events. Considering this, the analysis defines three signal regions: SRL, SRM, and SRH. The signal region SRL is designed to look for the parameter space with large $\Delta M (\Tilde{g}, \Tilde{\chi}_1^0)$\footnote{Because of this large mass difference, SRL can have many moderately hard jets resulting from the decay cascade. However, since the mass of the NLSP neutralino is low, the final state photons are rather soft, and the missing transverse momentum resulting from the final state gravitino is also relatively low.}. The signal region SRH, on the other hand, is designed to look for compressed spectra, i.e., small $\Delta M (\Tilde{g}, \Tilde{\chi}_1^0)$. Unlike SRL, SRH is characterized by a hard photon, a large missing transverse momentum, and few soft jets. Finally, the signal region SRM looks for the intermediate $\Delta M (\Tilde{g}, \Tilde{\chi}_1^0)$ scenario. Since the GGM parameter space studied by Ref \cite{ATLAS:2022ckd} encompasses high mass gluinos, the $H_T$ variable, defined as the scalar sum of the transverse momentum of the leading photon and signal jets, is expected to be high. Considering the high multiplicity of jets in the SRL and SRM signal regions we expect the sub-leading jets to be harder than those in the SM background events. Therefore, the variable $R_T^4$, defined as the ratio of the scalar sum of $p_T$ of the four leading jets to that of all signal region jets, is expected to take a relatively smaller value for the signal events. To remove events with significant contributions to $E_T^{\rm miss}$ from poorly reconstructed objects and instrumental sources, the analysis demands additional conditions on the angular separation between the $E_T^{\rm miss}$ and the jets/photons in the events. We summarise the three signal regions in Table \ref{tab:srsatlas}.\\

\begin{table}[htb!]
	\begin{center}
		\begin{tabular}{l|c|c|c}
			\toprule 
			 & SRL & SRM & SRH \\ 
			\toprule
			$N_{photons}$ & $\ge 1$& $\ge 1$& $\ge 1$\\
			
			$p_T^{leading-\gamma}$ &$> 145$ GeV &$ > 300$ GeV& $>400$ GeV \\
			
			$N_{leptons}$ &0 & 0& 0\\
			
			$N_{jets}$ &$\ge 5$ &$\ge 5$& $\ge 3$\\
			
			$\Delta \phi(jet,E_T^{miss})$ &$> 0.4$ &$> 0.4$&$> 0.4$\\
			
			$\Delta \phi(\gamma,E_T^{miss})$ &$> 0.4$ &$> 0.4$ &$> 0.4$\\
			
			$E_T^{miss}$ &$> 250$ GeV &$> 300$ GeV& $> 600$ GeV\\
			
			$H_T$ &$> 2000$ GeV &$> 1600$ GeV& $> 1600$ GeV\\
			
			$R_T^4$ &$< 0.90$ &$< 0.90$& - \\
            \bottomrule
            \bottomrule
            $\left\langle \epsilon \sigma\right\rangle^{95}_{obs}$(fb) &$0.034$ &$0.022$& $0.054$ \\

			\bottomrule
		\end{tabular} 
		\caption{\label{tab:srsatlas} Cuts used by the Atlas collaboration to define the three signal regions along with the model-independent 95 \% Confidence Level upper bound on the visible cross-section.}
	\end{center}
\end{table}

Dominant background contribution comes from processes like $t\Bar{t}\gamma$, W/Z$\gamma$, $\gamma$ + jets, $\gamma \gamma$, W$\gamma \gamma$, and Z$\gamma \gamma$. The absence of excess observed events over the estimated SM background events allows the analysis to set a 95 \% confidence level upper limit on the signal cross-section, which translates into lower bounds on the SUSY particle masses. The most stringent lower bound on the gluino mass was set at around 2.4 TeV for an NLSP mass of 1.3-1.4 TeV. Similarly, an overall lower limit is set on the gluino mass at 2.2 TeV for all NLSP masses except for $M_{\Tilde{\chi}_1^0}<150$ GeV and $M_{\Tilde{\chi}_1^0} >2050$ GeV.
\subsection{Revisiting mono-photon + $\slashed{E}_T$ search in Ref. \cite{ATLAS:2022ckd}}
\label{sec:limitatlasanalysis}
In this section, we will focus on one particular assumption of the ATLAS analysis \cite{ATLAS:2022ckd} that results in an overestimation of the lower bounds on the SUSY particle masses. Most experimental analyses dealing with BSM scenarios at LHC generally adhere to a simplified model that strategically omits certain couplings, often deemed irrelevant to the analysis. These assumptions render the model more manageable and phenomenologically accessible. However, when not handled carefully, such assumptions can also lead to a conclusion that might not be valid for a realistic scenario. One such assumption associated with the analysis of \cite{ATLAS:2022ckd} is that it only considers the gravitino decay mode of the lightest neutralino. In other words, it assumes that for the other neutralinos, charginos, and gluino, the gravitino decay mode is always suppressed compared to the cascade decay channel. Because of this assumption, the decay to the gravitino always occurs through the lightest neutralino, and the cascade decay always results in an energetic photon in the final state. However, as discussed in Section \ref{sec:decay}, there is a significant portion of the parameter space where this assumption does not hold, and the gravitino decay mode can dominate over the cascade decay channel. In such regions of parameter space, the pair-produced gluinos do not always result in a final state with photons, resulting in an event that falls outside the signal regions defined in the previous section. Under such circumstances, the constraints on the model parameter space become weaker.\\

For example, there are two regions of parameter space in particular where this problem becomes severe.
\begin{itemize}
    \item First is the parameter space leading to compressed spectra of the SUSY particle. As demonstrated in Figures \ref{fig:Brmupos} and \ref{fig:Brmuneg} (right panel), for both positive and negative $\mu$ scenarios, as the mass of the lightest neutralino approaches that of the gluino the branching ratio for the cascade decay channel becomes kinematically suppressed. There is a significant part of the parameter space where the cascade decay BR for gluino is even below ten percent.
    \item The second scenario that calls for attention is the case of negative $\mu$. As seen from the decay phase diagrams of $\Tilde{\chi}_2^0$ and $\Tilde{\chi}_1^{\pm}$ in Figure \ref{fig:Brmuneg} (left and middle panel), a significant portion of the parameter space favors direct decay of these particles into the gravitino than the cascade decay mode.
\end{itemize}

\begin{figure}[htb!]
	\centering
	\includegraphics[width=0.95\columnwidth]{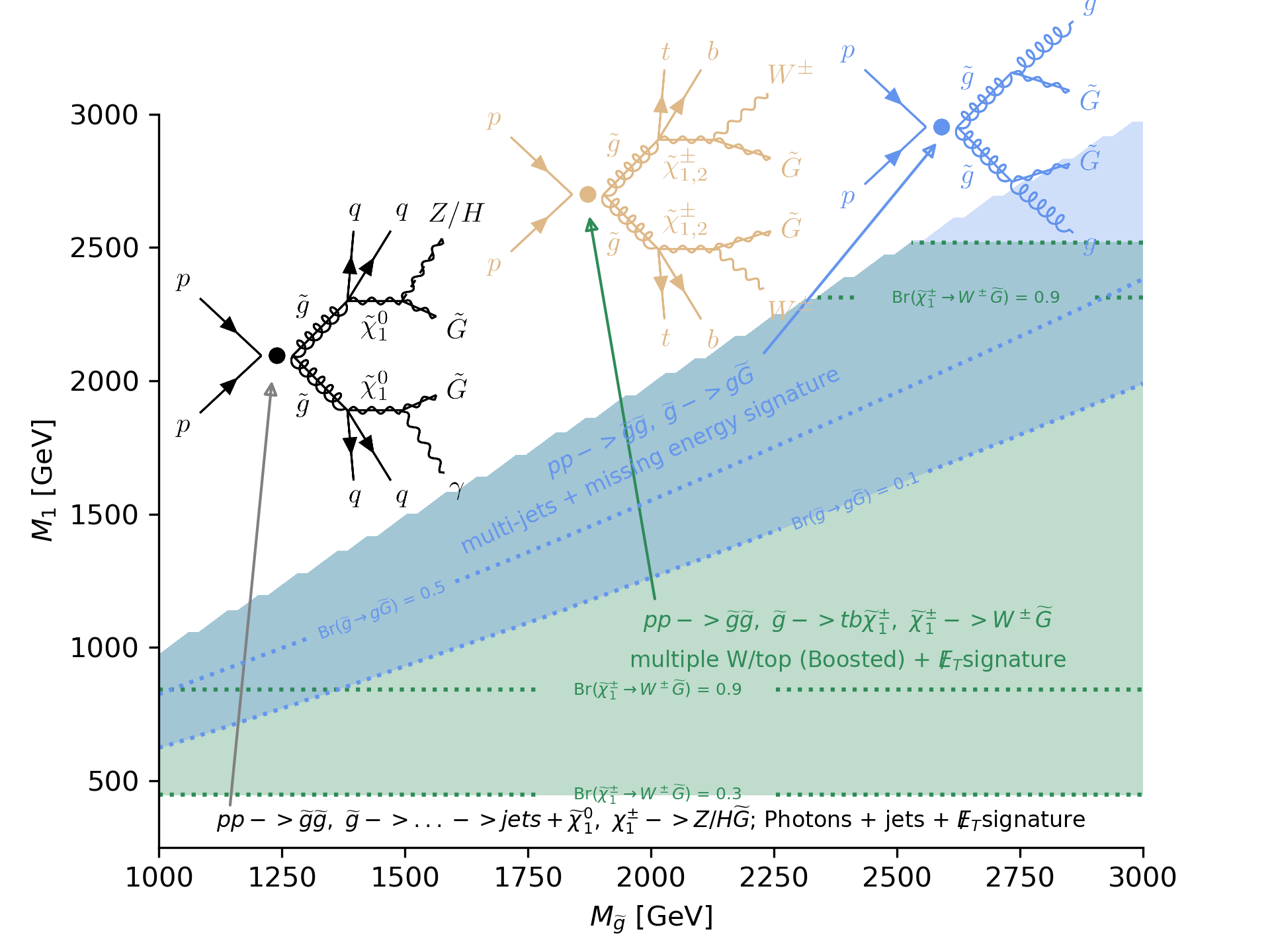}
	\caption{\label{fig:sig_topo_gammaH}Different regions of the $M_{\Tilde{g}}$ vs. $M_1$ plane give rise to different decay cascades and hence different final states for the pair-produced gluinos. These regions are schematically depicted in the figure, which illustrates the dominant decay channels and the resulting final states.
 }
\end{figure}

Different regions of $M_{\Tilde{g}}$ Vs $M_1$ plane giving rise to different decay cascade and hence different final states for the pair-produced gluinos are summarized in Figure \ref{fig:sig_topo_gammaH} (for the negative $\mu$ scenario) demonstrating the dominant decay modes in the different parts. The region in green presents the parameter space where the gravitino decay mode of the lightest chargino becomes gradually important. In the region between the two horizontal lines labeled BR($\Tilde{\chi}_1^{\pm} \rightarrow W^{\pm} \Tilde{G}$) = 0.9, the decay branching ratio of $\Tilde{\chi}_1^{\pm}$ in the gravitino mode exceeds 90 \%. As a result, though the gluino undergoes cascade decay, the decay to the LSP may proceed through the lightest chargino without producing any final state photon, instead, producing a hard $W$ boson. The region in blue shows the parameter space where the gravitino decay mode of the gluino becomes important. We have presented the region where the branching ratios in the gravitino decay mode exceed 10 \% and 50 \% by dotted lines. In this region, specifically for values of $M_{\Tilde{g}}$ close to $M_{\Tilde{\chi}_1^0}$ the gluino will prefer the direct decay channel to gravitino rather than the cascade decay mode giving rise to hard jets + $E_T^{\rm miss}$ final states. 

\subsection{Recasting the bounds on $m_{\tilde{g}}$--$m_{\tilde{\chi}_1^0}$ plane in the realistic GGM scenario}
\label{sec:analysis}
In this section, we recast the bounds on the $m_{\tilde{g}}$--$m_{\tilde{\chi}_1^0}$ plane from the ATLAS mono-photon search \cite{ATLAS:2022ckd} in the realistic-GGM scenario, considering all possible decays (both cascade decays and gravitino decays) for the SUSY particles. The technical details of our collider simulation and validation are given in the following.

\begin{enumerate}
    \item \textbf{Event Generation:} We use the default MSSM implementation of SARAH \cite{Staub:2013tta} and generate the SUSY particle spectrum in SPheno \cite{Porod:2011nf}. We generate gluino pair production with up to two additional partons in MG5\_AMC@NLO \cite{Alwall:2014hca}. For our analysis, we have used the NNLO$_{\rm{approx}}$ + NNLL gluino pair production cross section provided by the LHC SUSY Cross Section Working Group \cite{cern:susycrosssections}. Subsequent decays of the SUSY particles, showering, and hadronization are performed in Pythia8 \cite{Bierlich:2022pfr}. Note that the MSSM model file of SARAH does not contain the gravitino decay channels of the SUSY particles. To incorporate these decay channels into our analysis, we modify the decay tables of the MSSM particles after calculating the gravitino partial widths decay using the analytical expressions provided in Section~\ref{sec:decay}.
    
    \item \textbf{Object Reconstruction and Event Selection:} We use Delphes \cite{deFavereau:2013fsa} with the default ATLAS card to simulate the detector effects and reconstruct different physics objects like electrons, muons, photons, jets, and missing transverse energy. We closely followed the object reconstruction criteria discussed in Section~\ref{sec:atlasanalysis}. We also implemented the event selection criteria for the signal regions presented in Table~\ref{tab:srsatlas} to obtain the signal cross-sections resulting from gluino pair production in the realistic GGM scenario.
\end{enumerate}

\noindent \textbf{Validation:} For our final analysis, we divide the $M_{\tilde{g}}$ vs. $M_{\tilde{\chi}_1^0}$ plane into a grid and generate signal events for each of the grid cells. We calculate the signal events in three signal regions: SRL, SRM, and SRH, defined in Ref. \cite{ATLAS:2022ckd} and summarized in Table~\ref{tab:srsatlas}. We calculate the signal events in two different scenarios: the simplified GGM scenario considered in Ref. \cite{ATLAS:2022ckd}, where gravitino decay modes are considered only for NLSP-$\tilde{\chi}_1^0$ (the ATLAS scenario), and the realistic GGM scenario with allowed gravitino decays for all the SUSY particles. The comparison results derived for the former scenario with the ATLAS result in Ref. \cite{ATLAS:2022ckd} will provide the validation of our implementation of the model, event simulation, and event selection. To achieve comparable results, we calculate the (acceptance $\times$ efficiency) for each pair of ($M_{\tilde{g}}$, $M_{\tilde{\chi}_1^0}$) and compare them with the corresponding experimental results. We present our results for the ATLAS scenario in the left panels of Figures \ref{fig:eff_SRL_gammaH}, \ref{fig:eff_SRM_gammaH}, and \ref{fig:eff_SRH_gammaH} for the negative $\mu$ scenario and in the left panels of Figures \ref{fig:eff_SRL_gammaZ}, \ref{fig:eff_SRM_gammaZ}, and \ref{fig:eff_SRH_gammaZ} for the positive $\mu$ case for the signal regions SRL, SRM, and SRH, respectively. The color gradient in plots in Figs. \ref{fig:eff_SRL_gammaH}, \ref{fig:eff_SRM_gammaH}, \ref{fig:eff_SRH_gammaH}, \ref{fig:eff_SRL_gammaZ}, \ref{fig:eff_SRM_gammaZ}, and \ref{fig:eff_SRH_gammaZ} shows the (acceptance $\times$ efficiency) of different signal regions on the $M_{\tilde{g}}$--$M_{\tilde{\chi}_1^0}$ plane for the ATLAS scenario (left panels) and realistic-GGM scenario (right panels). For the few points on the $M_{\tilde{g}}$--$M_{\tilde{\chi}_1^0}$ plane in the left panels of the figures which correspond to the ATLAS scenario, we print our obtained values of (acceptance $\times$ efficiency) alongside the corresponding ATLAS results \cite{ATLAS:2022ckd}, which are in brackets. From this comparison, we can safely conclude that our results are very close to the corresponding ATLAS results \cite{ATLAS:2022ckd} for all the signal regions, proving the validity of our analysis.

\subsubsection{Final Results}
\label{sec:final}
\begin{figure}[htb!]
    \centering
    \includegraphics[width=0.495\columnwidth]{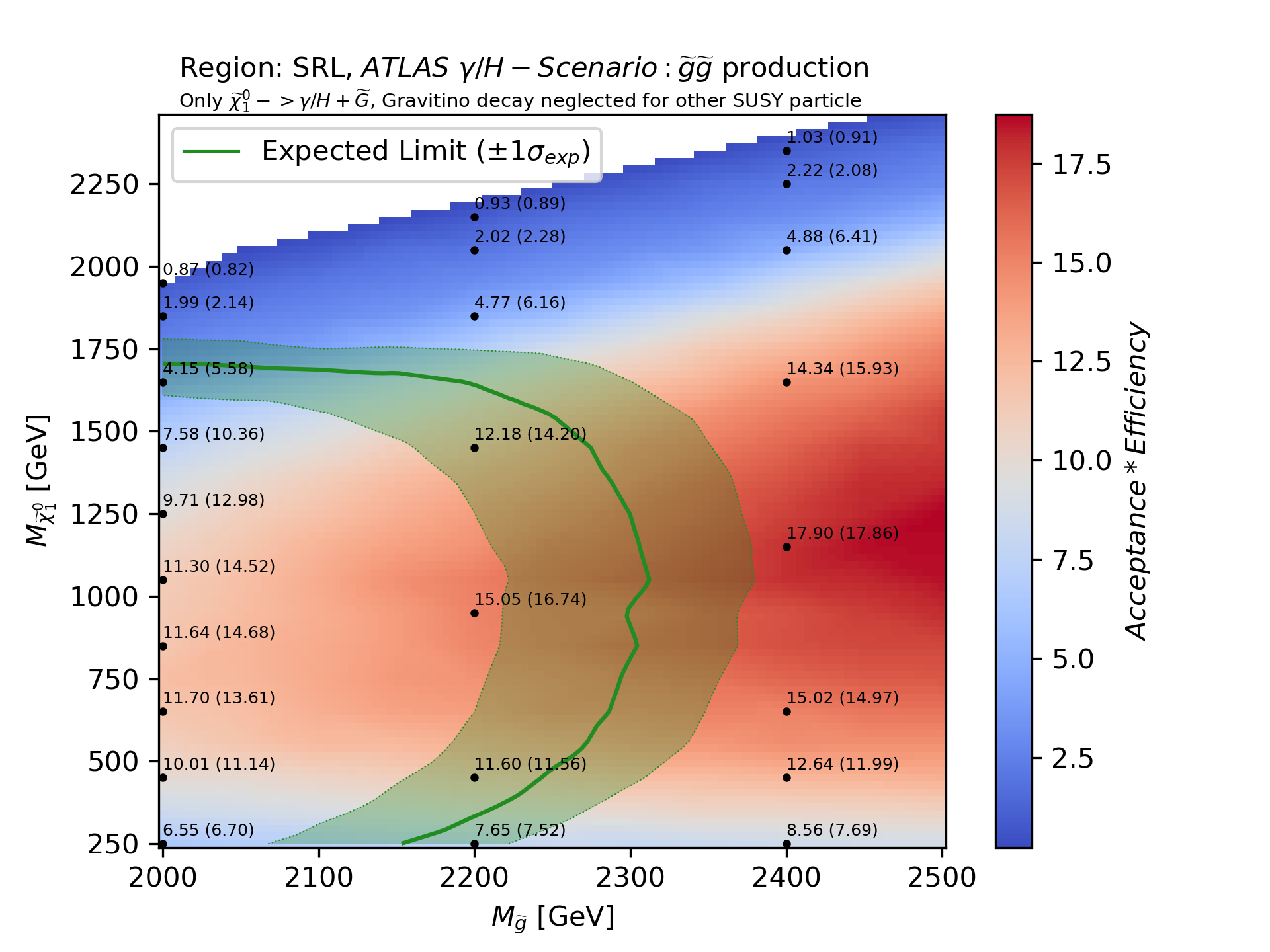}	
    \includegraphics[width=0.495\columnwidth]{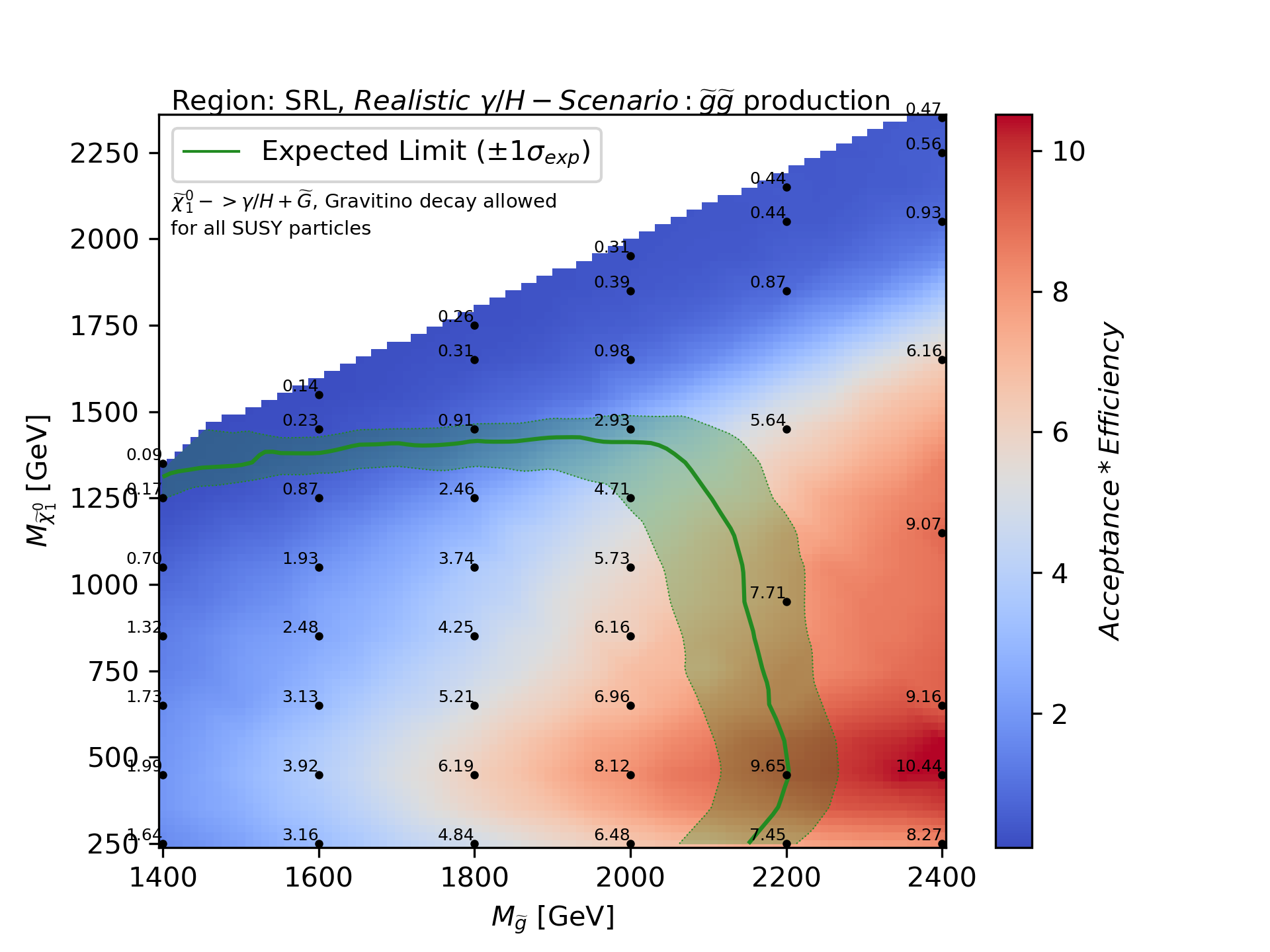}  
    \caption{\label{fig:eff_SRL_gammaH} The color gradient shows the acceptance $\times$ efficiency of signal region SRL for the ATLAS-considered scenario (left panel) and the realistic-GGM scenario (right panel) for $\mu < 0$. The green line corresponds to the 95\% CL expected limit, and the band represents the $\pm 1\sigma$ uncertainties on the expected limit due to both experimental and background-theory uncertainties. The numbers in brackets in the left panel correspond to the acceptance $\times$ efficiency obtained by the ATLAS collaboration in Ref. \cite{ATLAS:2022ckd} and are presented alongside the numbers obtained by us for validation.}
\end{figure}

\begin{figure}[htb!]
    \centering
    \includegraphics[width=0.495\columnwidth]{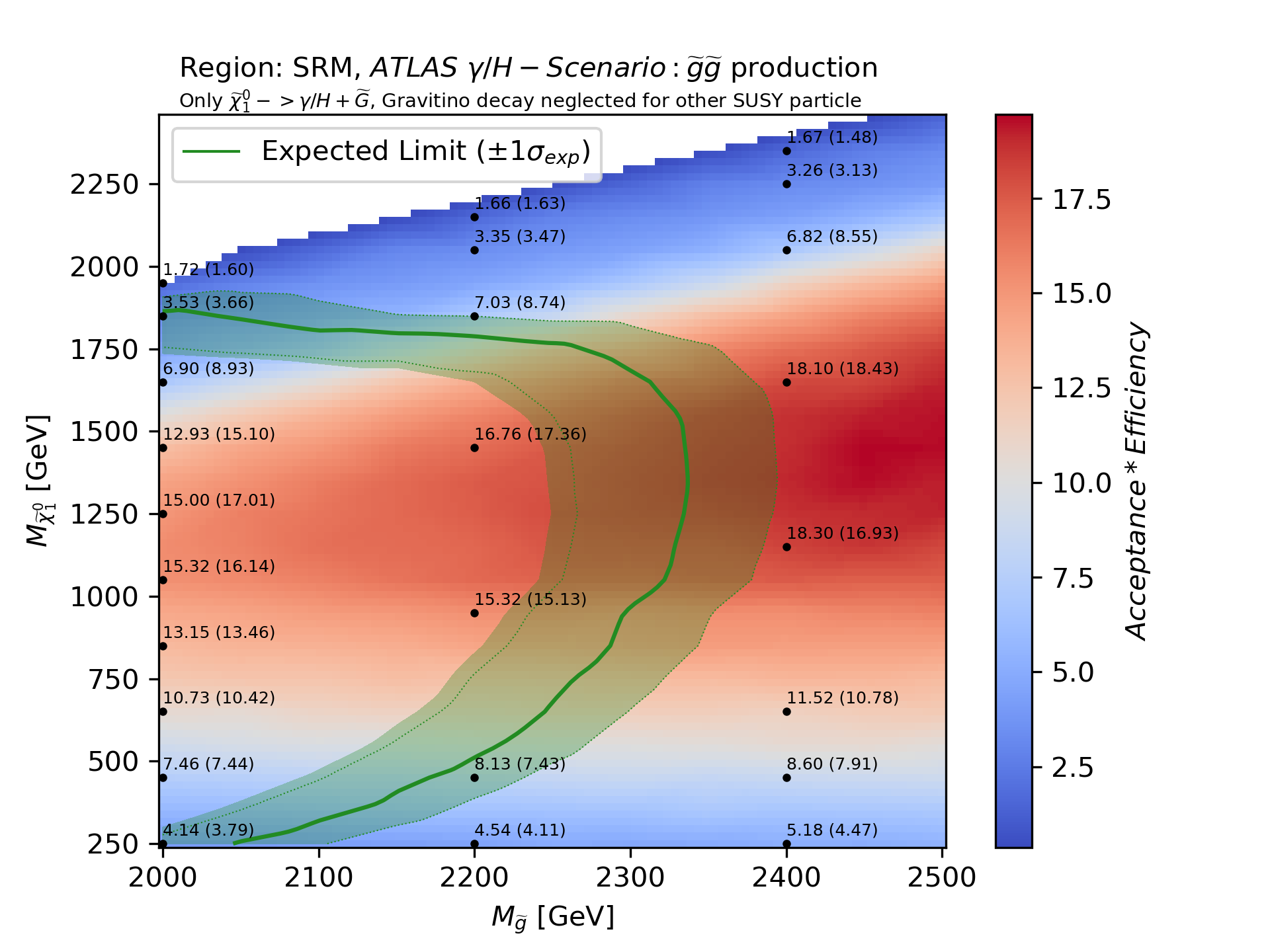}	
    \includegraphics[width=0.495\columnwidth]{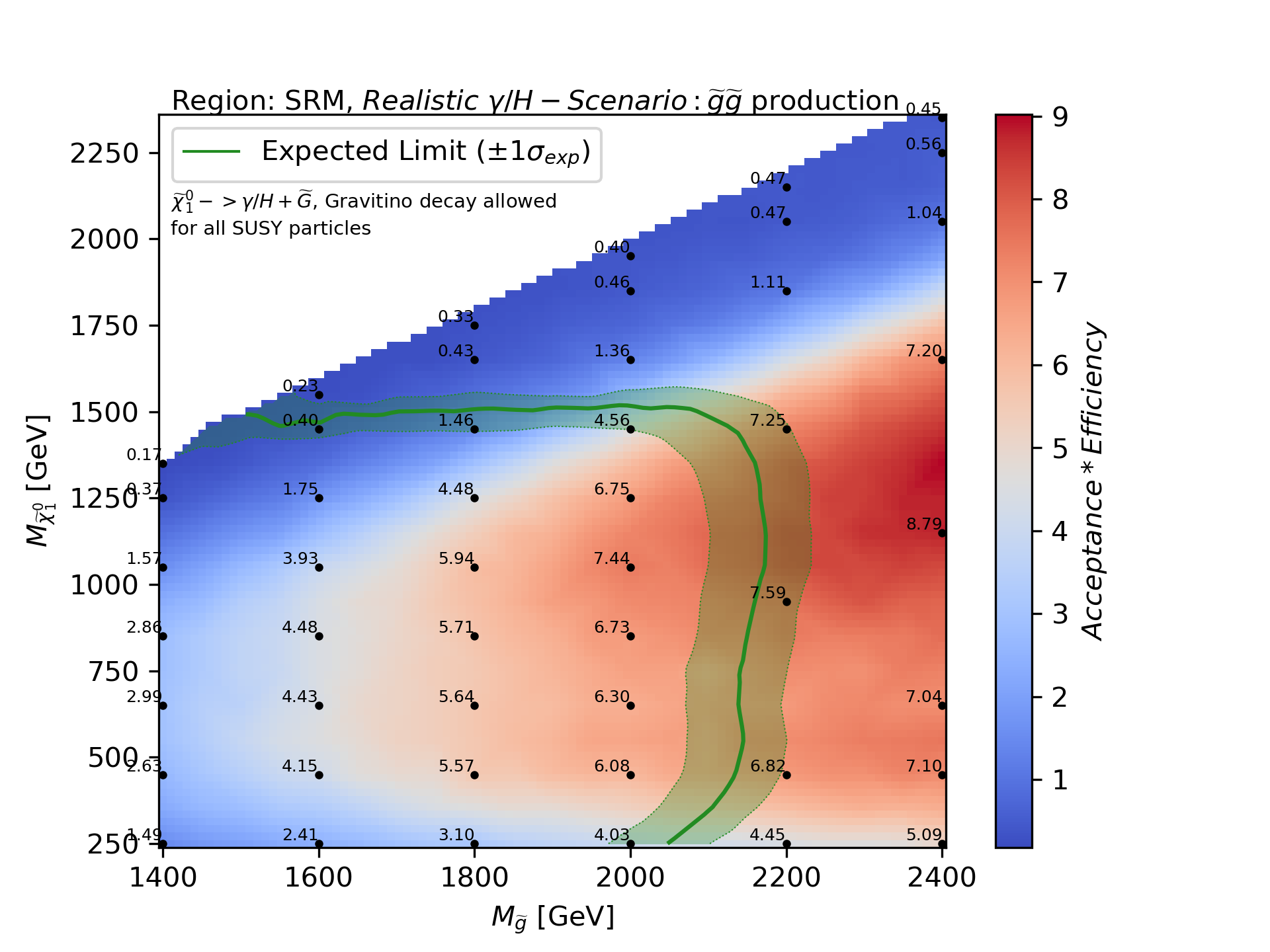}  
    \caption{\label{fig:eff_SRM_gammaH} Same as Fig. \ref{fig:eff_SRL_gammaH} for signal region SRM.}
\end{figure}

\begin{figure}[htb!]
    \centering
    \includegraphics[width=0.495\columnwidth]{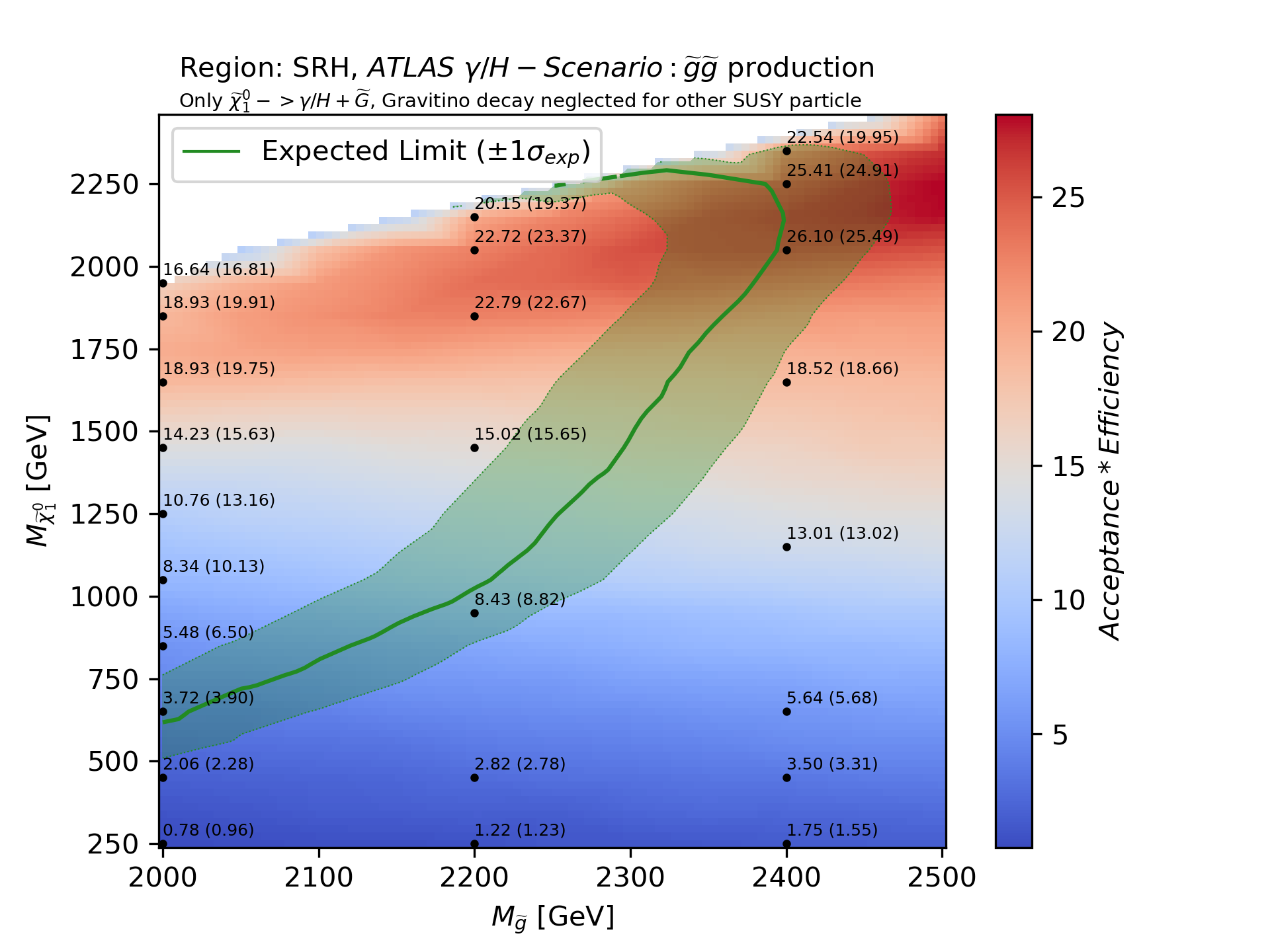}	
    \includegraphics[width=0.495\columnwidth]{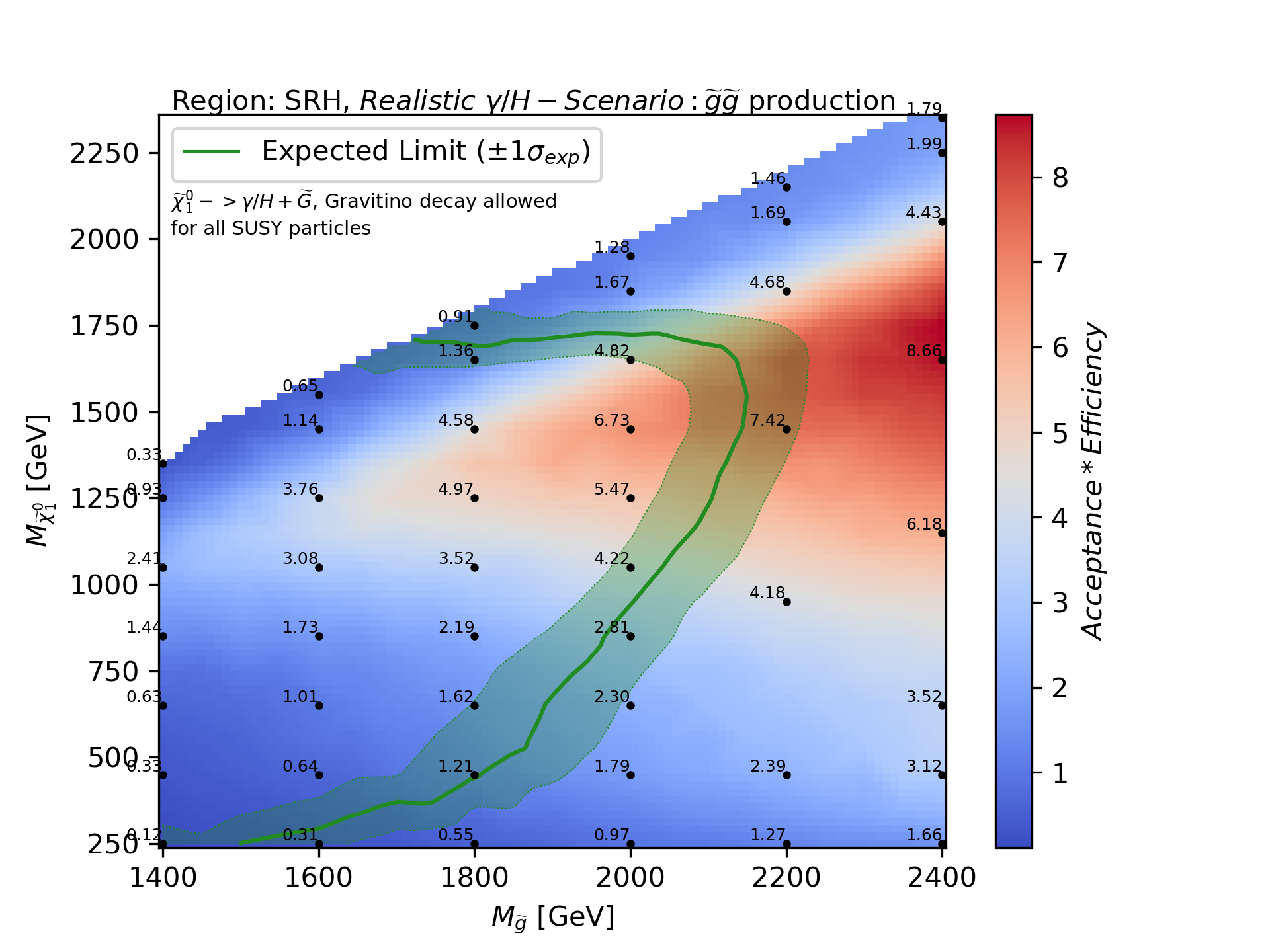}  
    \caption{\label{fig:eff_SRH_gammaH} Same as Fig. \ref{fig:eff_SRL_gammaH} for signal region SRH. }
\end{figure}

\begin{figure}[htb!]
    \centering
    \includegraphics[width=0.495\columnwidth]{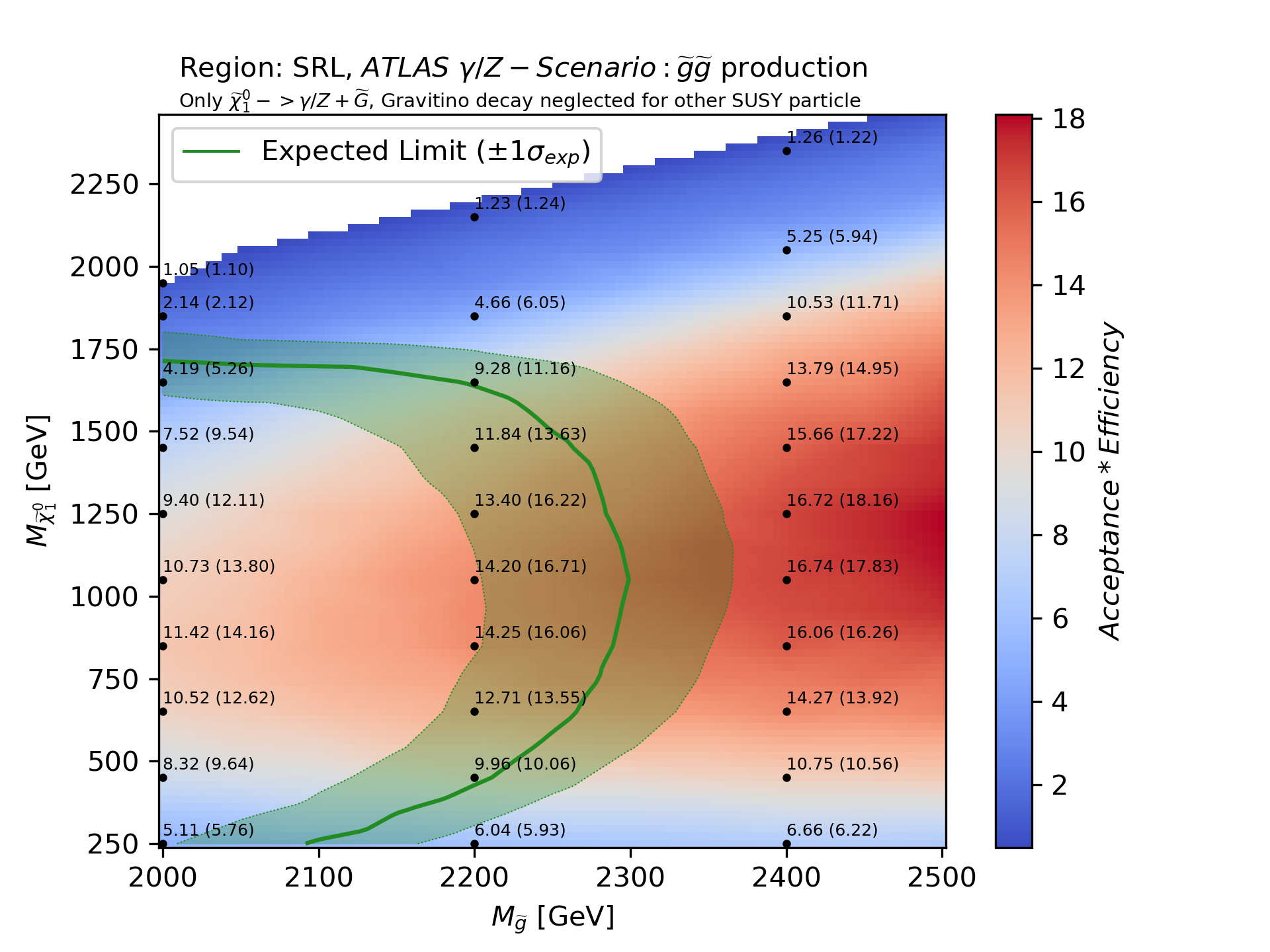}	
    \includegraphics[width=0.495\columnwidth]{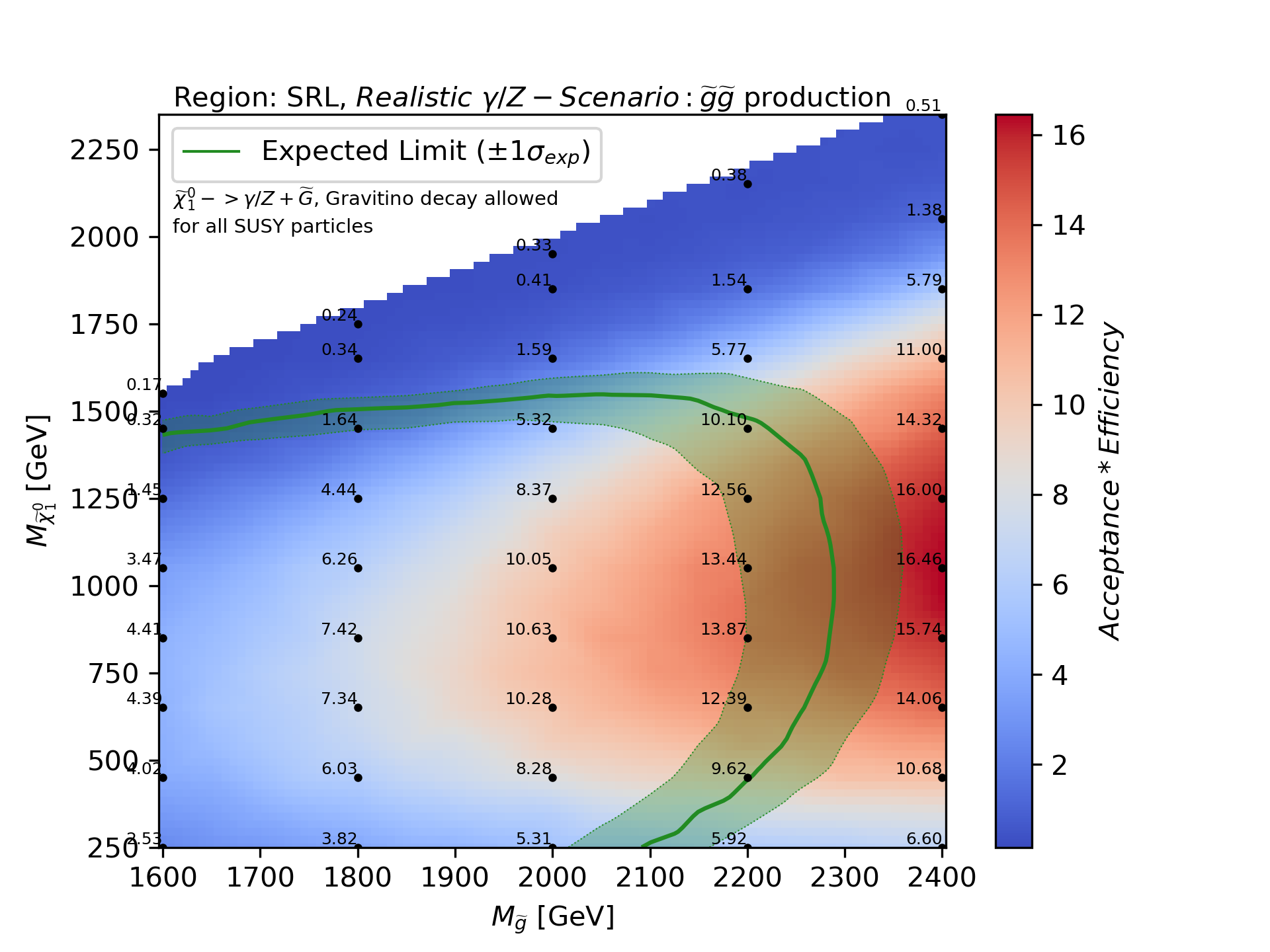}  
    \caption{\label{fig:eff_SRL_gammaZ} Same as Fig. \ref{fig:eff_SRL_gammaH} for $\mu > 0$. }
\end{figure}

\begin{figure}[htb!]
    \centering
    \includegraphics[width=0.495\columnwidth]{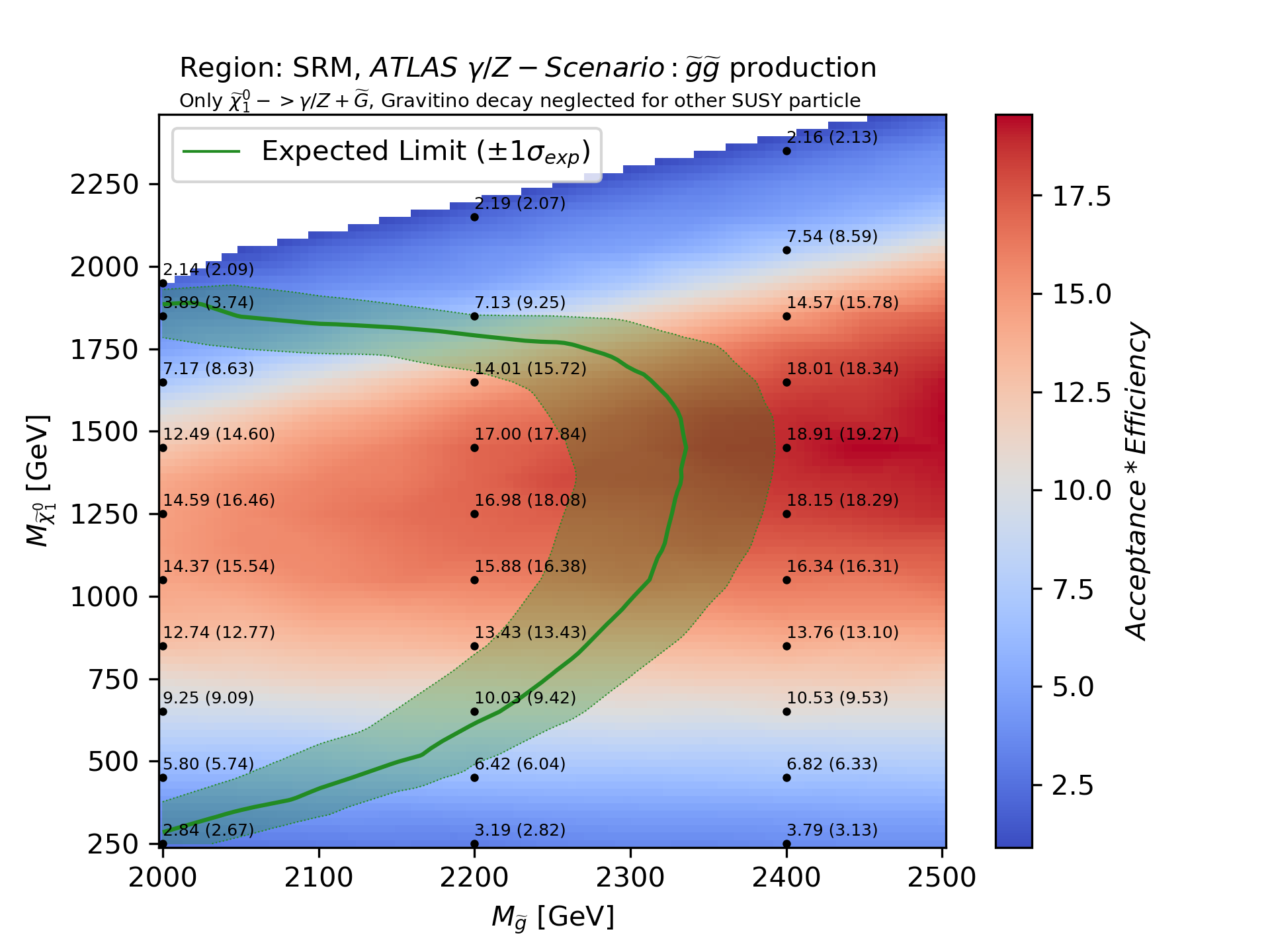}	
    \includegraphics[width=0.495\columnwidth]{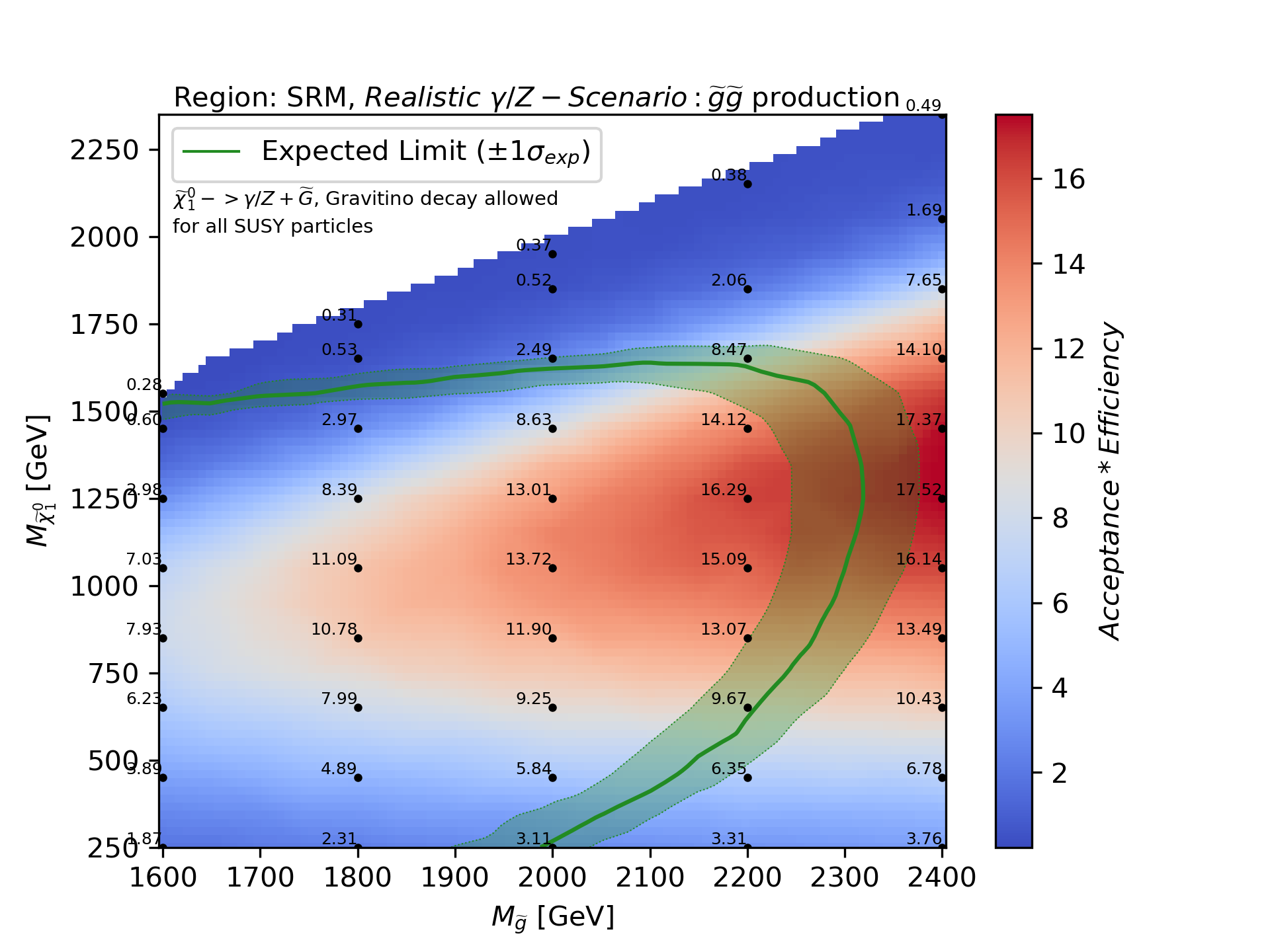}  
    \caption{\label{fig:eff_SRM_gammaZ} Same as Fig. \ref{fig:eff_SRM_gammaH} for $\mu > 0$. }
\end{figure}

\begin{figure}[htb!]
    \centering
    \includegraphics[width=0.495\columnwidth]{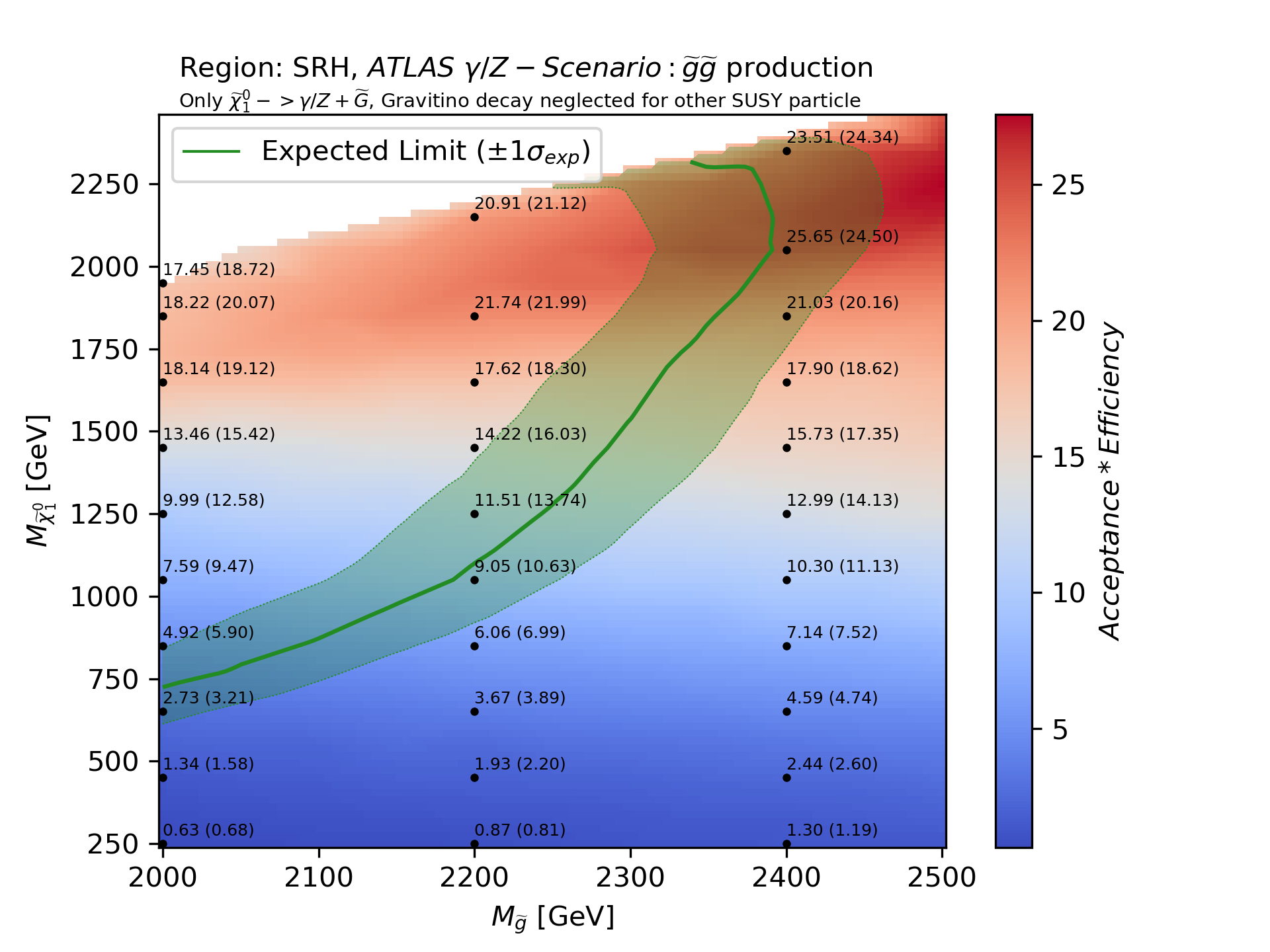}	
    \includegraphics[width=0.495\columnwidth]{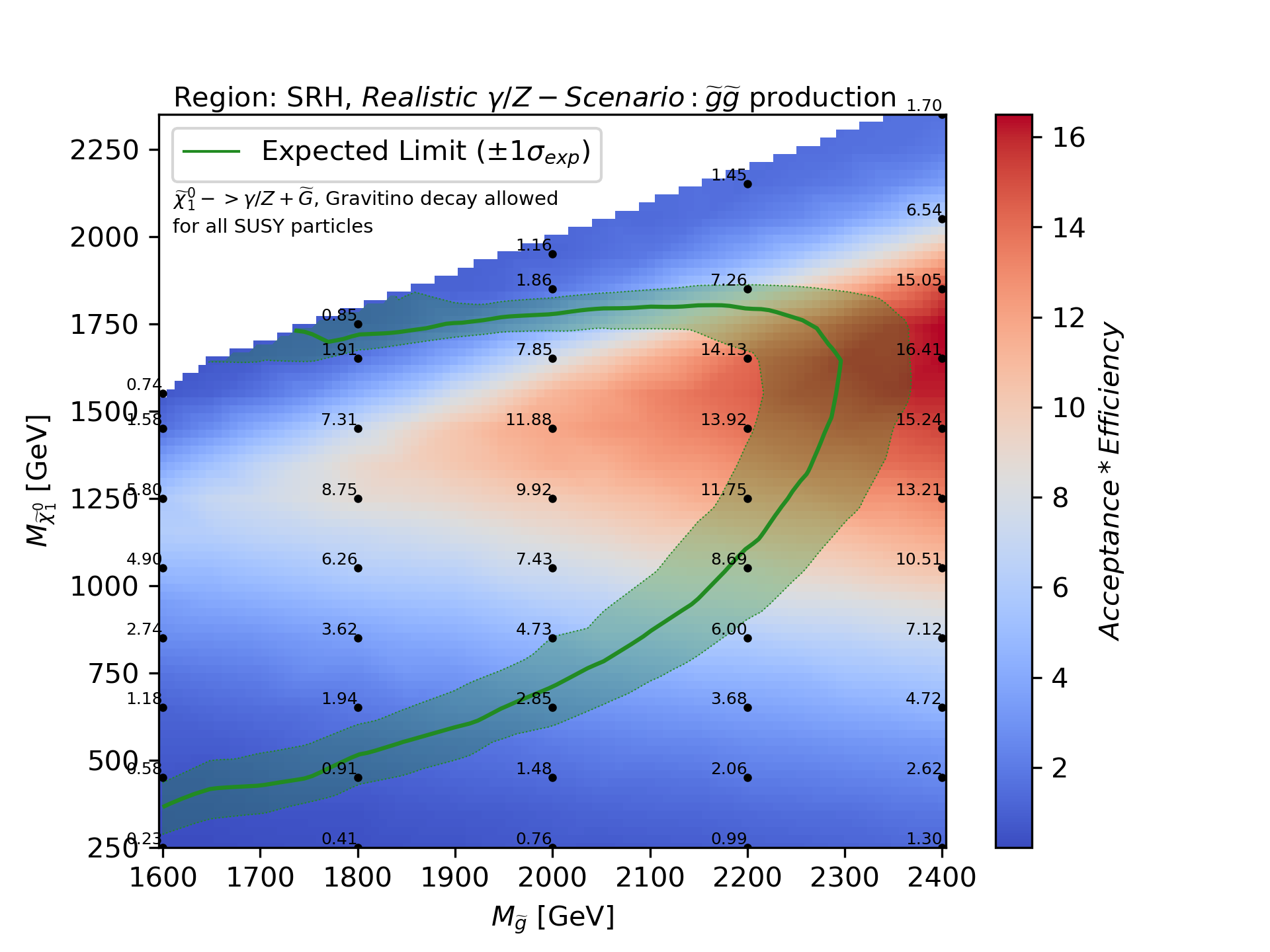}  
    \caption{\label{fig:eff_SRH_gammaZ} Same as Fig. \ref{fig:eff_SRH_gammaH} for $\mu > 0$. }
\end{figure}

To obtain the expected and observed bounds on the $M_{\tilde{g}}$--$M_{\tilde{\chi}_1^0}$ plane, we used the expected number of background events and the total number of observed events for the three signal regions from the ATLAS analysis \cite{ATLAS:2022ckd}. This choice enables us to compare our phenomenological results with those obtained by the experiment on an equal footing. To calculate the 95\% CL expected (observed) limits on the $M_{\tilde{g}}$--$M_{\tilde{\chi}_1^0}$ plane using SM background predictions (observed events) and the number of signal events in each signal region, we used the publicly available routine HistFitter \cite{Baak:2014wma}. In Figures \ref{fig:eff_SRL_gammaH} - \ref{fig:eff_SRH_gammaZ}, we present the expected limits for the different signal regions for both negative (Figures \ref{fig:eff_SRL_gammaH}, \ref{fig:eff_SRM_gammaH}, and \ref{fig:eff_SRH_gammaH}) and positive (Figures \ref{fig:eff_SRL_gammaZ}, \ref{fig:eff_SRM_gammaZ}, and \ref{fig:eff_SRH_gammaZ}) values of $\mu$. The green line in all these plots corresponds to the 95\% CL expected limit, and the band represents the $\pm 1\sigma$ uncertainties. While the plots on the left panels correspond to the ATLAS scenario \cite{ATLAS:2022ckd}, the expected limits of the right panel plots represent limits on $M_{\tilde{g}}$ and $M_{\tilde{\chi}_1^0}$ for a realistic GGM scenario with gravitino decays allowed for all the SUSY particles. As expected, the limits deviate significantly from the corresponding experimental results (left panel figures) after including all allowed gravitino decay modes. There are two regions where this deviation is particularly significant:

\begin{enumerate}
    \item \textbf{Quasi-degenerate gluino NLSP-neutralino region:} As discussed in Section \ref{sec:limitatlasanalysis}, the SRH signal region is designed to probe the parameter space resulting in a compressed gluino NLSP-neutralino spectra (small $\Delta M = M_{\tilde{g}}-M_{\tilde{\chi}_1^0}$). As can be seen in the right panels of Figures \ref{fig:Brmuneg} and \ref{fig:Brmupos}, the gravitino decay mode of the gluino dominates over the cascade decay channel for small $\Delta M$. As a result, in Figures \ref{fig:eff_SRH_gammaZ} and \ref{fig:eff_SRH_gammaH}, for both positive and negative $\mu$ scenarios, we obtain significantly relaxed bounds for the degenerate gluino NLSP-neutralino region of the parameter space.
    
    \item \textbf{Regions where gravitino decay modes for $\tilde{\chi}_2^0, \tilde{\chi}_3^0$, and $\tilde{\chi}_1^\pm$ dominate:} Figures \ref{fig:Brmuneg} (left and middle plots) show that there is a significant region of parameter space for negative $\mu$ where gravitino decays dominate for $\tilde{\chi}_2^0, \tilde{\chi}_3^0$, and $\tilde{\chi}_1^\pm$. When $\tilde{\chi}_2^0, \tilde{\chi}_3^0$, or $\tilde{\chi}_1^\pm$ appear in the decay cascade, they directly decay into a gravitino without decaying into the NLSP-neutralino, leading to a suppressed rate for the mono-photon final state studied in Ref. \cite{ATLAS:2022ckd}. This effect can be seen in Figures \ref{fig:eff_SRL_gammaH} and \ref{fig:eff_SRM_gammaH} as a suppressed reach for the gluino mass in the realistic scenario (right panel plots) compared to the ATLAS scenario (left panel plots).
\end{enumerate}


\begin{figure}[htb!]
    \centering
    \includegraphics[width=0.495\columnwidth]{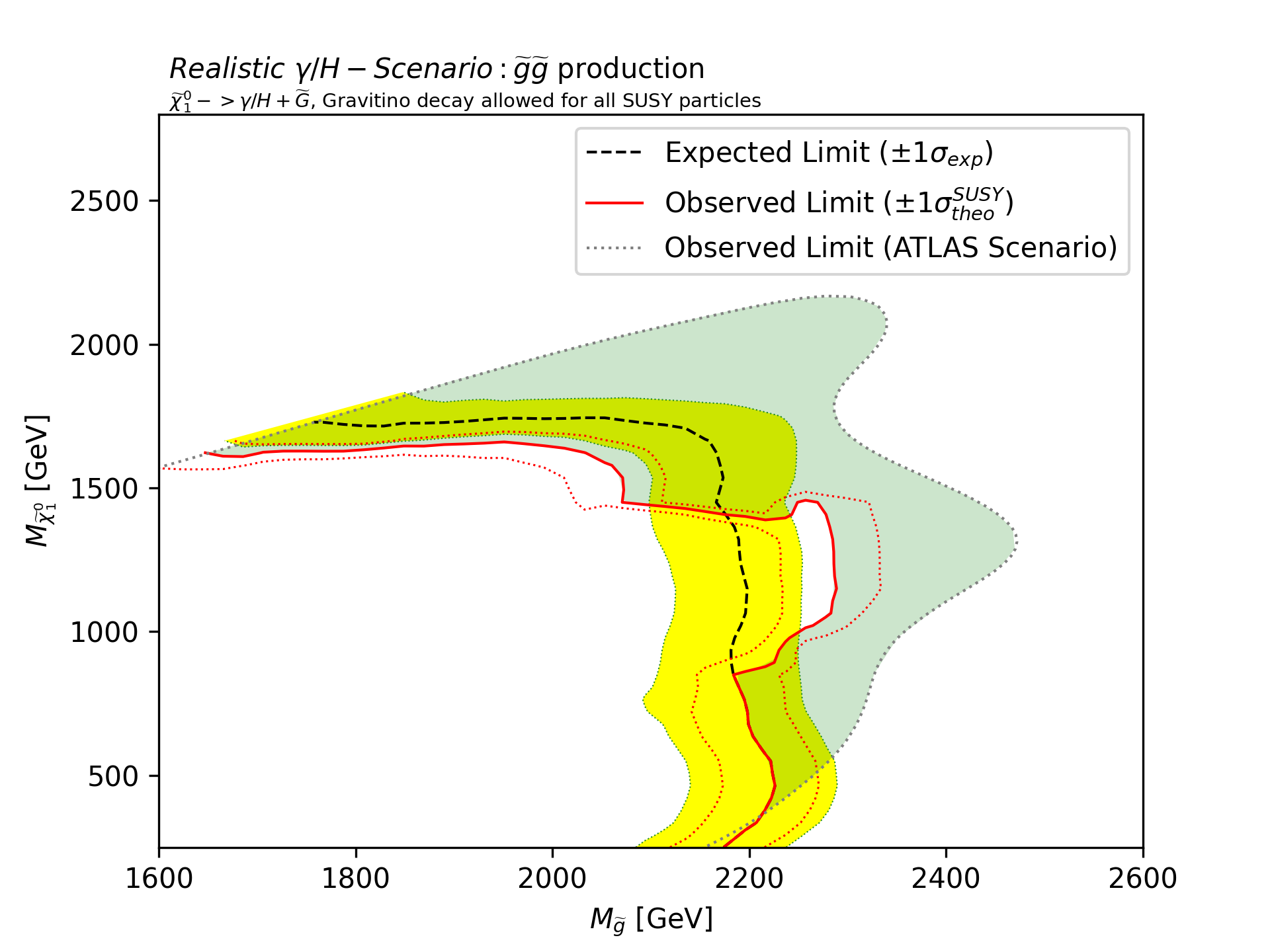}	
    \includegraphics[width=0.495\columnwidth]{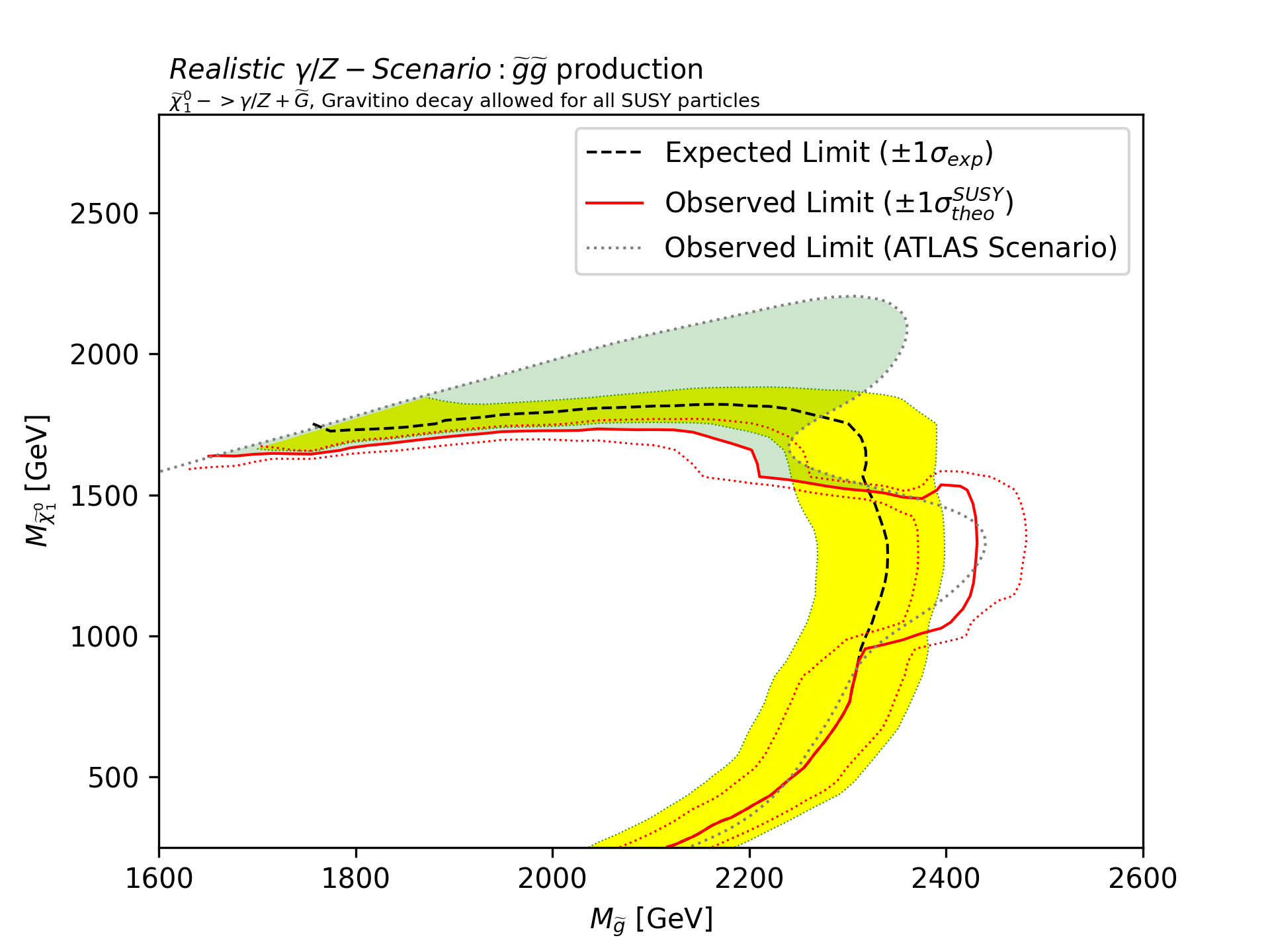}  
    \caption{\label{fig:bound} Observed and expected exclusion limits on the GGM model parameter space at 95\% CL, taking into account all possible decay modes of the SUSY particles for the negative (left) and positive (right) $\mu$ scenarios. For the sake of comparison, we have presented the ATLAS 95\% CL observed limit as a dotted line in both plots. The solid red line represents the 95 \% CL observed limit considering the nominal signal cross-section, with the dotted red lines representing the $\pm 1 \sigma$ error band due to theoretical uncertainties in the determination of the signal cross-section. The dashed black line represents the 95 \% CL expected limit, with the yellow band denoting the $\pm 1 \sigma$ exclusion band due to experimental and theoretical background uncertainties. The green-shaded region highlights the discrepancy between the ATLAS observed limit and our phenomenological observed limit.  }
\end{figure}

Finally, in Figure \ref{fig:bound}, we present the combined limits on the model parameter space in the $M_{\Tilde{g}}$ vs. $M_{\Tilde{\chi}_1^0}$ plane. The plot on the left (right) corresponds to the negative (positive) $\mu$ scenario. For the sake of comparison, we also present the observed limit from the experimental paper \cite{ATLAS:2022ckd} as a dotted line in both plots. To obtain our results,  for each value of ($M_{\Tilde{g}}$, $M_{\Tilde{\chi}_1^0}$), we pick the signal region with the best sensitivity. The dashed black line represents the 95 \% CL expected limit, with the yellow band denoting the $\pm 1 \sigma$ uncertainties. We use the numbers provided by the experimental analysis  \cite{ATLAS:2022ckd} for these uncertainties. Similarly, the solid red line represents the 95 \% CL observed limit considering the nominal signal cross-section, with the dotted red lines representing the $\pm 1 \sigma$ error band due to theoretical uncertainties (scale and PDF) in the determination of signal cross-section. For the signal cross-section along with the theoretical uncertainties, we use the values provided by the \textit{LHC SUSY Cross Section Working Group} \cite{cern:susycrosssections} for $\sqrt{s} = 13$ TeV LHC.  \\

As apparent from Figure \ref{fig:bound}, our observed limit differs from the ones obtained by the experimental collaboration. This discrepancy is less for the positive $\mu$ scenario and is mainly visible for values of $M_{\Tilde{\chi}_1^0}$ close to $M_{\Tilde{g}}$. On the other hand, our results differ significantly from the corresponding experimental ones for the negative $\mu$ scenario for reasons explained in the text. The experimental paper \cite{ATLAS:2022ckd} claimed a most stringent lower limit on the gluino mass of 2.4 TeV for both scenarios for values of $M_{\Tilde{\chi}_1^0}$ between 1.3 to 1.4 TeV (see the dotted line). As per our findings, this claim does not hold for the negative $\mu$ scenario. Our analysis suggests an updated (most stringent) lower limit of around 2.3 TeV for neutralino mass between 1.3 to 1.4 TeV. Furthermore, the experimental analysis \cite{ATLAS:2022ckd} put forward an overall lower limit on the gluino mass of 2.2 TeV for all neutralino masses except for $M_{\Tilde{\chi}_1^0} < $ 150 GeV and $M_{\Tilde{\chi}_1^0} > $ 2050 GeV. As per our analysis, this limit does not hold for values of $M_{\Tilde{\chi}_1^0}$ near $M_{\Tilde{g}}$. Quantitatively, the overall lower limit of 2.2 TeV only holds for $M_{\Tilde{\chi}_1^0} < $ 1500 (1300) GeV and $M_{\Tilde{\chi}_1^0} < $ 150 GeV for the positive (negative) $\mu$ scenario.

\section{Summary and Outlook}
\label{sec:summary}

Supersymmetry (SUSY) offers solutions to several fundamental issues in the Standard Model, such as the naturalness problem, gauge coupling unification, and a cosmologically viable candidate for dark matter. Nevertheless, SUSY must be broken at high energy scales to align with experimental observations. Various mechanisms for SUSY breaking have been proposed, including gravity mediation, anomaly mediation, and gauge mediation. This paper focuses on Gauge-Mediated Supersymmetry Breaking (GMSB) scenarios and their implications for collider searches, particularly within the context of the Large Hadron Collider (LHC).

The ATLAS collaboration has performed a mono-photon search with an integrated luminosity of 139 fb$^{-1}$ at a center-of-mass energy of 13 TeV to probe the signatures of simplified General Gauge Mediation (GGM) scenario. The original analysis relies on certain assumptions about the decay widths of SUSY particles into gravitinos, which may not be valid across the entire parameter space. The study presented in this paper reinterprets the ATLAS constraints on the gluino and NLSP-neutralino mass plane by considering a more comprehensive set of decay modes of SUSY particles in a realistic GGM model.


The reinterpretation uncovers significant discrepancies between the current experimental limits in the framework of a simplified GGM scenario with unrealistic assumptions and a realistic GGM scenario with complex decay cascades considering gravitino decay modes for all the relevant SUSY particles. For instance, the lower experimental limit on the gluino mass is not applicable in certain regions where the gluino mass is close to the neutralino mass. The study suggests an updated lower limit of approximately 2.3 TeV on gluino mass for neutralino masses between 1.3 to 1.4 TeV, which contrasts with the previously reported limit of 2.4 TeV by the experimental analysis. Additionally, the overall lower limit on the gluino mass of 2.2 TeV is only valid for neutralino masses below 1500 GeV in the positive $\mu$ scenario and below 1300 GeV in the negative $\mu$ scenario. In certain regions of $M_{\Tilde{g}} - M_{\Tilde{\chi}_1^0}$ plane, the deviation of the ATLAS limit could be as large as 300 GeV.


The discrepancy (illustrated by the green-shaded region in Figure \ref{fig:bound}) between the ATLAS limit in the context of the simplified GGM scenario and our reinterpretation of ATLAS data within the realistic GGM scenario is attributed to photon-deficiency in the final state in this specific region of the parameter space. A deeper understanding of this photon-deficiency in the final state can be gained from Figure \ref{fig:sig_topo_gammaH}, which illustrates the dominant decay modes of SUSY particles across various values of $M_{\Tilde{g}}$ and $M_1$. In these regions, while decays of the SUSY particles to the photonic final states are suppressed, alternative decay channels involving top quarks and W/Z bosons become important. Given the heavy mass of the SUSY particles, the resulting SM particles are expected to be highly boosted, meaning their decay products at the LHC can often be reconstructed within a single large-radius fat jet. Consequently, a search strategy targeting these boosted particles could effectively probe this region. However, such an analysis is beyond the scope of our current study and will be addressed in future work.

\acknowledgments KG acknowledges the Helsinki Institute of Physics for hosting their visit, which facilitated the completion of part of this work.

\bibliographystyle{JHEP}
\bibliography{v0}

\end{document}